\documentclass[aip,pop,twocolumn,preprintnumbers,
amsmath,amssymb,floatfix,nofootinbib,reprint]
{revtex4-1}

\usepackage{amsmath}
\usepackage{amsfonts}
\usepackage{amssymb}
\usepackage{graphicx}%
\usepackage{dcolumn}%
\usepackage{bm}%
\usepackage{color}
\usepackage{multirow}
\usepackage{hyperref}
\usepackage[caption=false, font=small]{subfig}
\usepackage[utf8]{inputenc}

\DeclareFontFamily{U}{BOONDOX-calo}{\skewchar\font=45 }
\DeclareFontShape{U}{BOONDOX-calo}{m}{n}{
  <-> s*[1.05] BOONDOX-r-calo}{}
\DeclareFontShape{U}{BOONDOX-calo}{b}{n}{
  <-> s*[1.05] BOONDOX-b-calo}{}
\DeclareMathAlphabet{\mathcalboondox}{U}{BOONDOX-calo}{m}{n}
\SetMathAlphabet{\mathcalboondox}{bold}{U}{BOONDOX-calo}{b}{n}
\DeclareMathAlphabet{\mathbcalboondox}{U}{BOONDOX-calo}{b}{n}

\newcommand{\e}{\text{e}}

\renewcommand{\i}{\text{i}}

\newcommand{\be}{\begin{equation}}
\newcommand{\ee}{\end{equation}}
\newcommand{\beqa}{\begin{align}}
\newcommand{\eeqa}{\end{align}}
\newcommand{\beqan}{\begin{align*}}
\newcommand{\eeqan}{\end{align*}}

\def\ket#1{$| #1 \rangle$}
\def\ketm#1{| #1 \rangle}

\def\me#1#2#3{$\langle #1 | #2 | #3 \rangle$}
\def\mem#1#2#3{\langle #1 | #2 | #3 \rangle}

\def\threejm#1#2#3#4#5#6{\left( \begin{matrix} #1 & #2 & #3  \\
                                               #4 & #5 & #6  \end{matrix} \right)}
\begin{document}

 \urlstyle{same}
%
%
%
%

%
%

\title{Direct and secondary nuclear excitation with x-ray free-electron lasers}

%
%

\author{Jonas \surname{Gunst}}
\affiliation{Max-Planck-Institut f\"ur Kernphysik, Saupfercheckweg 1, D-69117 Heidelberg, Germany}

\author{Yuanbin \surname{Wu}}
\email{yuanbin.wu@mpi-hd.mpg.de}
\affiliation{Max-Planck-Institut f\"ur Kernphysik, Saupfercheckweg 1, D-69117 Heidelberg, Germany}

\author{Naveen \surname{Kumar}}
\affiliation{Max-Planck-Institut f\"ur Kernphysik, Saupfercheckweg 1, D-69117 Heidelberg, Germany}

\author{Christoph H. \surname{Keitel}}
\affiliation{Max-Planck-Institut f\"ur Kernphysik, Saupfercheckweg 1, D-69117 Heidelberg, Germany}

\author{Adriana \surname{P\'alffy}}
\email{Palffy@mpi-hd.mpg.de}
\affiliation{Max-Planck-Institut f\"ur Kernphysik, Saupfercheckweg 1, D-69117 Heidelberg, Germany}


\date{\today}

%
%
%
%
%
%
%
\begin{abstract}

The direct and secondary nuclear excitation produced by an x-ray free electron laser when interacting with a solid-state nuclear target is investigated theoretically. When driven at the resonance energy, the x-ray free electron laser can produce direct photoexcitation. However, the dominant process in that interaction is the photoelectric effect producing a cold and very dense plasma in which also secondary processes such as nuclear excitation by electron capture may occur. We develop a realistic theoretical model to quantify the temporal dynamics of the plasma and the magnitude of the secondary excitation therein. Numerical results show that depending on the nuclear transition energy and the temperature and charge states reached in the plasma, secondary nuclear excitation by electron capture may dominate the direct photoexcitation by several orders of magnitude, as it is the case for the 4.8 keV transition from the isomeric state of $^{93}$Mo, or it can be negligible, as it is the case for the 14.4 keV M\"ossbauer transition in $^{57}\mathrm{Fe}$. These findings are most relevant for future nuclear quantum optics experiments at  x-ray free electron laser facilities.

\end{abstract}


\maketitle

\section{Introduction}
The recently operational X-ray Free Electron Laser (XFEL) facilities \cite{LCLS-web,SACLA-web} have considerably enhanced the focus on x-ray physics, with particular emphasis on x-ray quantum optics \cite{Adams2013.JoMO} and the extension of concepts used in atomic physics towards the high-frequency radiation domain \cite{}. A peculiar circumstance is that x-rays are no longer resonant to valence-electron transitions in atoms. Instead, the corresponding resonant systems are either inner shell electron transitions in (highly) charged ions \cite{Young2010.N,Kanter2011.PRL,Rohringer2012}, or transitions in atomic nuclei \cite{Buervenich2006,Rohlsberger2012.N}. Nuclear quantum optics featuring the interaction of x-ray light with Mössbauer nuclei in the few keV transition energy range has gained considerable momentum, both theoretically \cite{Kocharovskaya1999,Buervenich2006,Palffy2009,Liao2012a,heeg2013x,Brinke2013,Liao2014,heeg2015y} and experimentally \cite{Shvydko1996.PRL,Rohlsberger2010.S,Rohlsberger2012.N,heeg2013vacuum,Vagizov2014,Slowlight2015,Fano2015}.
Interestingly, these experiments operate with less than one
resonant x-ray photon per pulse on average \cite{Gerdau1999.HI} since they make use of synchrotron sources with brilliance values about eight orders of magnitude lower than the ones at the XFEL. This raises the question what qualitative and quantitative differences are to be expected in nuclear quantum optics experiments to be performed in the near future at XFEL sources?
Considering the fundamental laser-matter interaction processes involved, this is also a new situation compared to photonuclear studies with petawatt optical lasers \cite{KenL2000.PRL,Cowan2000.PRL,Gibbon2005.Book,Spohr2008.NJP,Mourou2011.S} like, for instance, the currently built LFEX \cite{LFEX} in Osaka or ELI-NP \cite{ELI-NP-web} in Bucharest.

\begin{figure}
\centering
\includegraphics[width=1.0\linewidth]{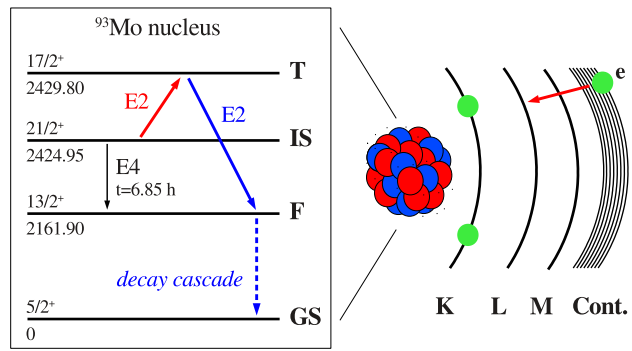}%
\caption{(color online)  Schematic illustration of nuclear excitation by electron capture \cite{Jonas2013.PRL}. A free electron (right-hand side) recombines into a highly charged ion with the simultaneous excitation of the nucleus (simplified level scheme on the left-hand side). In this example, the $^{93}$Mo nucleus is initially in a metastable state. The
$^{93\mathrm{m}}$Mo excitation is induced by NEEC into the $L$ shell (right-hand side) with subsequent decay to the nuclear ground state (long blue solid and dashed arrows in the left panel). The nuclear levels are labeled with their total angular momentum, parity  and energy (in keV). Reproduced with permission from Phys. Rev. Lett. 112, 082501 (2014). Copyright 2014 American Physical Society.
\label{fig:neec}}
\end{figure}

While so far in all x-ray-nucleus experiments the electronic response only acted as background, the increase of the electric field strength leads to drastic changes in the interaction between photons and electrons which may additionally influence the nuclear excitation. For  photon energies of approx.~10 keV and middle-range atomic number $Z$ materials, the photoeffect dominates the electronic processes. Due to the unique interaction between high intensity x--ray pulses and matter \cite{Young2010.N,Hau-Riege.book}, new matter states like cold, high--density plasmas can be reached \cite{Lee2003.JOSAB,Vinko2012.N}. In such environments secondary nuclear processes from the coupling  to the atomic shell are rendered possible by the presence of free electrons and atomic vacancies. 
For instance, in the resonant process of nuclear excitation by electron capture (NEEC) \cite{Palffy2006.PRA},  a free electron is captured into a bound state simultaneously transferring the excess energy to the nucleus, as illustrated schematically in Fig.~\ref{fig:neec}.
While the related process nuclear excitation by electron transition (NEET) has been observed, e.g., in experiments with ${}^{197}$Au \cite{Kishimoto-NEET}, an experimental evidence of NEEC remained elusive so far, mainly due to strong radiative background processes.

Secondary nuclear processes in the XFEL-produced plasma open new channels of nuclear excitation. Depending on the goal of the nuclear quantum optics experiment, this might be of advantage, for instance if the maximization of nuclear excitation is desired, or of disadvantage, if  coherence-based enhancement of nonlinear interaction between x-rays and nuclei is envisaged. In the latter case the interactions in the plasma would lead to strong decoherence rates.
In a recent work \cite{Jonas2013.PRL} we have shown theoretically that the  secondary NEEC occurring in the plasma environment created by the XFEL pulse 
shining on a bulk nuclear solid-state target  can exceed by orders of magnitude the direct resonant photoexcitation. The concrete example studied was the case of photoexcitation starting from the isomeric, long-lived nuclear excited state of $^{93}$Mo at approx.~2.5 MeV energy. The XFEL-induced 4.85 keV excitation may open in this case the possibility to retrieve the energy stored in the metastable state in a process known as  isomer triggering or isomer depletion  \cite{Walker1999.N,Aprahamian2005.NP,Belic1999.PRL,Collins1999.PRL,Belic2002.PRC,Carroll2004.L,Palffy2007.PRL}. In this first comparison between direct and secondary nuclear excitation processes, the temporal evolution of the plasma states was neglected, i.e., NEEC was assumed to take place in the same plasma conditions for a period of 100 ps. In reality, the plasma undergoes Coulomb explosion during this time and we may expect that only a reduced fraction of the assumed 100 ps is effectively providing the advantageous parameters for secondary nuclear excitation in the plasma. This critical approximation in Ref.~\cite{Jonas2013.PRL}  calls therefore for further investigation.

The purpose of the present work is twofold. First, we would like to provide a more reliable theoretical prediction on the secondary nuclear excitation  by taking into account the time-dependent dynamics of the plasma. We do so by parameterizing the plasma expansion 
by means of a hydrodynamical model \cite{KrainovS2002,DitmireDRFP1996} in the quasi-neutral limit and including atomic processes in the plasma with the help of the  population kinetics model implemented in the FLYCHK code \cite{FLYCHK}. Employing this improved model, we provide here a more accurate comparison of the direct and secondary nuclear excitation in the isomer triggering process investigated in Ref.~\cite{Jonas2013.PRL}. The second motivation is to provide reliable theoretical predictions for direct and secondary nuclear excitation in the case of the 14.4 keV M\"ossbauer transition in $^{57}\mathrm{Fe}$. That nucleus has been the ``working horse'' of the nuclear forward scattering community at synchrotron sources \cite{Shvydko1996.PRL,Roehlsberger2004} and it is also the present candidate for x-ray quantum optics using nuclear transitions in thin film x-ray planar cavities \cite{Rohlsberger2010.S, Rohlsberger2012.N,heeg2013vacuum}. Many of the present studies at synchrotron sources have extensions envisaged with the XFEL as source of stronger, nonlinear nuclear excitation. For these studies it is vital to know whether also 
additional plasma effects may play an important role for the nuclear excitation or for the sought-for coherence effects.

Our results show that, although for the case of isomer triggering in  $^{93}$Mo, secondary nuclear excitation via NEEC remains dominant by several orders of magnitude, this is not at all the case for $^{57}\mathrm{Fe}$, where the secondary excitation can be safely neglected. We identify here criteria related to the nuclear transition energy, the atomic structure and plasma conditions that can be used to identify whether for a particular nuclear species, the secondary processes can be dominant compared to the direct photoexcitation channel. This knowledge is then applied to the present nuclear transition candidates starting from stable or metastable ground states and energies in the operation range of the XFEL.  These results are most relevant for the layout of future nuclear quantum optics experiments at the XFEL.

The paper is structured as follows. In Section \ref{theory} the theoretical formalisms used to describe photexcitation and NEEC in the plasma environment are briefly reviewed. The hydrodynamical expansion model used for the plasma is introduced here. Section \ref{results} presents our numerical results on direct and secondary nuclear excitation. A detailed analysis of the plasma condition is included. This section concludes with a comparison between the two excitation channels and a general set of qualitative criteria to identify the dominant nuclear excitation channel. The main conclusions of the paper are summarized in the last section.

%

\section{Theory \label{theory}}

The theoretical formalisms used to address the direct photoexcitation and the secondary NEEC process in the XFEL-produced cold plasma are briefly reviewed in the following. The plasma dynamics is described by means of a hydrodynamical model. The effects of the latter on the secondary excitation channel are quantified. Atomic units $\hbar=m_{\rm{e}}=e=4\pi\varepsilon_0=1$ are used throughout Secs.~\ref{theophoto} and \ref{sec:theory.neec}.

\subsection{Direct light-nucleus interaction \label{theophoto}}

The natural and most convenient way to describe the direct  interaction of the laser electromagnetic field with  the two-level nuclear system is via the well-established density matrix formalism. The detailed description of this treatment has been presented elsewhere  \cite{Palffy2008.PRC,Palffy2008.JoMO} and we restrict ourselves here to only few relevant expressions.
The time evolution of the density matrix $\rho$ is determined by the master equation \cite{ScullyZubairy}
\begin{equation}
i \frac{\partial}{\partial t} \rho=[H_0+H_{\rm{I}},\rho]+\mathcal{L}\rho\ ,
\label{eq:master}
\end{equation}
where the so-called Lindblad operator $\mathcal{L}$ includes decoherent relaxation processes of the system like spontaneous decay or dephasing of the laser field. The system Hamiltonian can be separated into the unperturbed part $H_0$ and the interaction part $H_{\rm{I}}$.
The dynamics of the system is then governed by the interaction Hamiltonian $H_{\rm{I}}$ given by
\begin{equation}
H_{\rm{I}}=-\frac{1}{c}\int d^3r \,  \bm{j}_{\rm{n}}(\bm{r})\cdot \bm{A}(\bm{r},t)\ ,
\label{H_int}
\end{equation}
where $c$ stands for the speed of light and $\bm{j}_{\rm{n}}$ represents the nuclear current density. The vector potential of the electromagnetic field denoted by $\bm{A}$ is described classically. Making use of the multipole expansion and applying the long-wavelength approximation it can be shown that the interaction Hamiltonian $H_{\rm{I}}$ is directly proportional to the electric and magnetic multipole moments $\mathcalboondox{Q}_{LM}$ and $\mathcalboondox{M}_{LM}$ which are defined by
\begin{eqnarray}
\mathcalboondox{Q}_{LM}&=&\int d^3r\, r^{L} Y_{LM}(\theta,\varphi)\rho_{\rm{n}}(\bm{r})\ ,
\nonumber \\
\mathcalboondox{M}_{LM}&=&-\frac{\i}{c}\sqrt{\frac{L}{L+1}}\int d^3r \, r^{L}\bm{Y}^{M}_{LL}(\theta,\varphi)\cdot\bm{j}_{\rm{n}}(\bm{r})\ .
\label{eq:multipoles} 
\end{eqnarray}
Here, $L$ denotes the total photon angular momentum and $M$ its projection,
$\rho_{\rm{n}}$ represents the nuclear charge density and $\bm{Y}^{M}_{LL}$ stand for the vector spherical harmonics $\bm{Y}^{M}_{JL}$ with $J=L$ \cite{Edmonds}.

The calculation of the nuclear interaction matrix elements then reduces to the evaluation of \me{I_{\rm{e}} M_{\rm{e}}}{\mathcalboondox{Q}_{LM}}{I_{\rm{g}} M_{\rm{g}}} and \me{I_{\rm{e}} M_{\rm{e}}}{\mathcalboondox{M}_{LM}}{I_{\rm{g}} M_{\rm{g}}}, where the ground (excited) nuclear state wave functions are labeled by the total nuclear spin $I_{\rm{g}}(I_{\rm{e}})$ and its projection $M_{\rm{g}}(M_{\rm{e}})$.  Utilizing the Wigner-Eckart theorem \cite{Edmonds} the matrix elements can be related to the reduced transition probabilities $\mathcal{B}$ which are defined as \cite{RingSchuck}
\be
\mathcal{B} (\mathcal{E}(\mathcal{M})L,I_{\rm{g}}\to  I_{\rm{e}})=\frac{1}{2I_{\rm{g}}+1}\big|\langle I_{\rm{e}}\|\mathcalboondox{Q}_{L}(\mathcalboondox{M}_{L})\|I_{\rm{g}}\rangle \big|^2\ .
\ee
Here, $\mathcal{E}$ stands for electric and $\mathcal{M}$ for magnetic parity transitions, respectively.
Since often experimental values are available for the reduced transition probabilities $\mathcal{B}$, one can avoid the limitations of theoretical nuclear models.
For an electric (magnetic) transition of fixed multipolarity $L$,  the interaction matrix element reads \cite{Palffy2008.PRC}
\begin{align}
&\langle I_{\rm{e}} M_{\rm{e}}|H_{\rm{I}}|I_{\rm{g}} M_{\rm{g}}\rangle
\nonumber \\
\propto&\, E_k \e^{-\i\omega_kt}\sqrt{2\pi} \sqrt{\frac{L+1}{L}}C\big(I_{\rm{e}}\ I_{\rm{g}}\ L;M_{\rm{e}}\ -\!M_{\rm{g}}\ M_{\rm{e}}\!-\!M_{\rm{g}}\big)
\nonumber \\
&\times \frac{k^{L-1}}{(2L+1)!!} \sqrt{2I_{\rm{g}}+1} \sqrt{\mathcal{B}({\mathcal{E}(\mathcal{M})}L,I_{\rm{g}}\to I_{\rm{e}}) }\ ,
\label{matrHI}
\end{align}
where $k$ denotes the wave number, $\omega_k$ the angular frequency and $E_k$  the electric field amplitude of the laser, respectively, and  $C(j_1\ j_2\ j_3; m_1\ m_2\ m_3)$ stands for the relevant Clebsch-Gordan coefficient.

The natural nuclear transition width $\Gamma_0$ that enters the Lindblad operator in Eq.~(\ref{eq:master}) is usually composed of two parts: (i) the radiative decay where the nuclear excitation energy is released in form of a photon, and (ii) internal conversion (IC) which accelerates a bound electron into the continuum due to the nuclear deexcitation. 
Typically, $\Gamma_0$ is in the range of $10^{-5}$ to $10^{-10}$~eV for low-lying nuclear transitions which are accessible with the XFEL. The large discrepancy between the nuclear width $\Gamma_0$ and the XFEL bandwidths (currently in the order of several eV) is one of the main limiting factors for the direct laser-nucleus interaction. However, if solid-state targets are considered, there can be a third decay channel, the collective or coherent decay, which may strongly enhance the nuclear transition width and hence the laser-induced nuclear excitation. This collective decay relies on the fact that in solid-state targets composed of M\"{o}ssbauer nuclei, photons can be absorbed and re-emitted without recoil. Due to the recoilless scattering it is impossible to distinguish which nucleus has been excited such that a collective excitation  also known as nuclear exciton is created \cite{NFS-bible}. The formation of the exciton requires that after excitation, each nucleus decays  to the initial ground state. Otherwise,  processes like decay by IC, nuclear recoil or spin-flip which provide (be it even only in principle) ``which-nucleus'' information prevent the formation of the collective effect. 
 The collective excitation predominantly decays by emitting a photon into the forward direction with an increased  decay rate \cite{NFS-bible}. By considering a very short incoming laser pulse, approximated as a $\delta$ function  in time, the time dependence of the scattered intensity immediately after nuclear excitation can be approximated as \cite{NFS-bible,Shvydko1999.PRB,Junker2012.NJP}
\begin{equation}
I(t) \propto e^{-(\xi+1)\Gamma_0 t}\ ,
\end{equation}
where $\xi=\frac{1}{4}\sigma_{\rm{R}} n_0 L$ stands for the dimensionless thickness parameter with $n_0$ denoting the number density of nuclei in a sample of length $L$ and $\sigma_{\rm{R}}$ the resonant nuclear cross section.
The faster decay rate expression $(\xi+1)\Gamma_0$ is only valid in relation to the formation of the nuclear exciton. For instance, while the excitonic state is typically populated in nuclear forward scattering experiments with $^{57}\mathrm{Fe}$, this cannot be the case for $^{93}$Mo. The reason for the latter is that due to the isomer triggering nuclear level scheme (see Fig.~\ref{fig:neec}), the final nuclear state after the decay $F$ does not coincide with the initial state $IS$. This allows at least in principle a way to discern which nuclei have been excited, and prohibits collective effects.

\subsection{NEEC: microscopic description  \label{sec:theory.neec}}
A detailed presentation of the NEEC theoretical formalism can be found in Refs.~\cite{Palffy2006.PRA,Palffy2007.PRA}. For completion, here we reproduce some relevant expressions that are used in our later numerical analysis.
The initial state for NEEC  describes  the system  composed of the nucleus in its ground or a metastable excited state, the bound electrons indexed by the quantum numbers $\alpha_0$, the total electronic angular momentum quantum number $J_0$ and its projection $\mu_0$, and additionally a free electron with asymptotic momentum $\mathbf{p}$ and free electron spin projection $m_{\rm{s}}$ \cite{Palffy2006.PRA},
\begin{equation}
\ketm{\psi_{\rm{i}}} = \ketm{ I_{\rm{i}} M_{\rm{i}}, (\alpha_0 J_0 \mu_0,\mathbf{p} m_{\rm{s}})}\ .
\label{eq:initial.state}
\end{equation}
In the process of NEEC, the free electron is captured by simultaneously exciting the nucleus to some state \ket{I_{\rm{d}} M_{\rm{d}}}. The electronic capture level is denoted by the quantum numbers $\alpha_{\rm{d}}$, $J_{\rm{d}}$ and $\mu_{\rm{d}}$. With that, the intermediate state can be written as
\begin{equation}
\ketm{\psi_{\rm{d}}} = \ketm{ I_{\rm{d}} M_{\rm{d}}, \alpha_{\rm{d}} J_{\rm{d}} \mu_{\rm{d}}}\ .
\label{eq:intermediate.state}
\end{equation}

For the description of the electron-nucleus interaction we adopt the Coulomb gauge expression
\begin{equation}
H_{\rm{en}} = \int d^3 r_{\rm{n}} \, \frac{\rho_{\rm{n}}(\mathbf{r}_{\rm{n}})}{\left|\mathbf{r}_{\rm{n}} - \mathbf{r}_{\rm{e}} \right|}\ ,
\label{eq:en.hamiltonian}
\end{equation}
where  $\mathbf{r}_{\rm{n,e}}$ stand for the nuclear and electronic spatial coordinates, respectively. Important to note is that the attractive Coulomb interaction in Eq.~(\ref{eq:en.hamiltonian}) can only mediate transitions of electric multipolarity. The magnetic electron-nucleus interaction is instead transferred via the exchange of a virtual photon between the nuclear and electronic currents \cite{Palffy2006.PRA}. The corresponding interaction Hamiltonian $H_{\rm{m}}$ is obtained by performing a perturbation expansion of the system's transition operator and making use of the Feshbach projection method \cite{Palffy2006.PRA}. According to Ref.~\cite{Palffy2006.PRA} the magnetic Hamiltonian can be written as
\begin{equation}
H_{\rm{m}} = -\frac{1}{c} \boldsymbol{\alpha} \cdot \int d^3 r_{\rm{n}} \, \frac{\mathbf{j}_{\rm{n}}(\mathbf{r}_{\rm{n}})}{\left| \mathbf{r}_{\rm{e}} - \mathbf{r}_{\rm{n}} \right|}\ ,
\label{eq:magn.hamiltonian}
\end{equation}
where $\boldsymbol{\alpha}$ is the vector composed of the three Dirac matrices $\alpha_{\rm{x}}$, $\alpha_{\rm{y}}$ and $\alpha_{\rm{z}}$.

For the calculation of the NEEC transition rate we assume that the projection quantum numbers of the initial and final states are not resolved in the experiment. This is accomplished by averaging over $M_{\rm{i}}$, $\mu_0$ and $m_{\rm{s}}$ and summing over $M_{\rm{d}}$ and $\mu_{\rm{d}}$. Furthermore, we integrate over the solid angle $\Omega_p$ of the incoming electrons. For the general case of both electric and/or magnetic transitions,  the NEEC transition rate reads \cite{Palffy2006.PRA,Palffy2007.PRA}
\begin{eqnarray}
Y^{\rm{i} \rightarrow \rm{d}}_{\rm{neec}} &=& \frac{2 \pi}{2 (2 I_{\rm{i}}+1) (2 J_0+1)} \sum_{M_{\rm{i}} m_{\rm{s}} \mu_0} \sum_{M_{\rm{d}} \mu_{\rm{d}}} \int d\Omega_p \nonumber \\ 
&\times&  \left| \mem{\psi_{\rm{d}}}{H_{\rm{en}} + H_{\rm{m}}}{\psi_{\rm{i}}} \right|^2 \rho_{\rm{i}}\ ,
\label{eq:neec.rate}
\end{eqnarray}
where $\rho_{\rm{i}}$ stands for the initial density of free electron states. The final result for the NEEC rate defined above for capture into ions with closed shells ($J_0=0$), a certain multipolarity $L$ and a transition of electric parity then reads
\begin{align}
 \label{rate}
{Y^{(\mathcal{E} L)}_{\rm{neec}}}=&\frac{4\pi^2\rho_{\rm{i}}}{(2L+1)^2}R_{\rm{n}}^{-2(L+2)} \mathcal{B}(\mathcal{E}L,I_{\rm{i}}\to I_{\rm{d}}) (2J_{\rm{d}}+1)
\nonumber \\ &\times
\sum_\kappa \left|R^{(\mathcal{E})}_{L,\kappa_{\rm{d}},\kappa}\right|^2\
C\big(J_{\rm{d}}\ L\ j;\frac{1}{2}\ 0\ \frac{1}{2}\big)^2\ ,
\end{align}
where $j$ is the total angular momentum of the continuum electron, and $\kappa$ is the Dirac angular momentum quantum number related to $j$ via $j=|\kappa|-\frac{1}{2}$. The index $\kappa_{\rm d}$ denotes the Dirac angular momentum quantum number of the bound electron after the capture. 
In the case of a magnetic transition with multipolarity $L$ we obtain
\begin{align}
{Y_{\rm{neec}}^{(\mathcal{M} L)}}=&\frac{4\pi^2\rho_{\rm{i}}}{L^2(2L+1)^2} \mathcal{B}(\mathcal{M}L,I_{\rm{i}}\to  I_{\rm{d}}) (2J_{\rm{d}}+1) \sum_{\kappa}
\nonumber \\&\times (2j+1)
(\kappa_{\rm{d}}+\kappa)^2 \threejm{J_{\rm{d}}}{j}{L}{\frac{1}{2}}{-\frac{1}{2}}{0}^2 \left|R^{(\mathcal{M})}_{L,\kappa_{\rm{d}},\kappa}\right|^2\ .
\end{align}
In the expressions above, $R_{\rm{n}}$ stands for the nuclear radius and the occurring radial integrals $R^{(\mathcal{E})}_{L,\kappa_{\rm{d}},\kappa}$ and $R^{(\mathcal{M})}_{L,\kappa_{\rm{d}},\kappa}$ require a numerical approach to be solved. Their definitions can be found in Refs.~\cite{Palffy2006.PRA,Palffy2007.PRA}.

The NEEC cross section is then given by
\begin{equation}
\sigma^{\rm{i} \to \rm{d} }_{\rm{neec}}(E) = \frac{2\pi^2}{p^2} 
 Y_{\rm{neec}}^{\rm{i} \to \rm{d}} L_{\rm{d}}(E-E_{\rm{d}})
\label{eq:sigma-neec}
\end{equation}
where $p$ is the absolute value of the initial free electron momentum $\bm{p}$. The function $L_{\rm{d}}$ represents the well-known normalized Lorentz profile occurring in resonant systems defined as
\begin{equation}
L_{\rm{d}}(E-E_{\rm{d}}) = \frac{\Gamma_{\rm{d}} / 2\pi}{(E-E_{\rm{d}})^2 + \frac{1}{4} \Gamma_{\rm{d}}^2}\ ,
\label{lorentz}
\end{equation}
with $E_{\rm{d}}$ and $\Gamma_{\rm{d}}$ being the energy and the natural width of the resonant state, respectively. For arbitrary capture levels $\psi_{\rm{d}}$ the width $\Gamma_{\rm{d}}$ is composed of an electronic and a nuclear part, $\Gamma_{\rm{d}} = \Gamma_{\rm{d}}^{\rm{el}} + \Gamma_{\rm{d}}^{\rm{nucl}}$.

\subsection{NEEC reaction rates in the plasma environment\label{neecinplasma}}
In a plasma, free electrons with different kinetic energies are available. 
NEEC on the other hand is a resonant process where a free electron is captured by an ion with the simultaneous excitation of the nucleus. The 
resonance bandwidth is determined by the Lorentz profile (\ref{lorentz}). 
Since the kinetic energy of free electrons in a plasma is distributed over a wide range, many resonant NEEC channels may exist. In the following we will shortly describe how such a situation can be handled theoretically in terms of reaction rates.

As introduced in Sec.~\ref{sec:theory.neec}, the initial and intermediate states are given by Eq.~(\ref{eq:initial.state}) and Eq.~(\ref{eq:intermediate.state}), respectively. In order to restrict the number of possible initial configurations, for a lower-limit estimate,
we consider in the following only NEEC into ions which are in their electronic ground state. In this case, the initial electronic configuration $\alpha_{\rm 0}$ is uniquely identified by the charge state number $q$ before electron capture. 
In the isolated resonance approximation, the total NEEC reaction rate in the plasma can be written as a summation over all charge states $q$ and all capture channels $\alpha_{\rm{d}}$,
\begin{equation}
\lambda_{\rm{neec}} = \sum_q \sum_{\alpha_{\rm{d}}} P_q \lambda^{q,\alpha_{\rm{d}}}_{\rm{neec}}\ ,
\label{eq:neec-plasma.total}
\end{equation}
where the partial NEEC rate into the capture level $\alpha_{\rm{d}}$ of an ion in the charge state $q$ is given by
\begin{equation}
\lambda^{q,\alpha_{\rm{d}}}_{\rm{neec}} = \int dE \, \sigma_{\rm{neec}}^{\rm{i} \to \rm{d}}(E) \phi_{\rm{e}}(E)\ .
\label{eq:neec-plasma.partial}
\end{equation}
The single-resonance  cross sections $\sigma^{\rm{i} \rightarrow \rm{d}}_{\rm{neec}}$ are defined in Eq.~(\ref{eq:sigma-neec}). 
The dependence on $q$ is hidden here in the NEEC resonance energy $E_{\rm{d}}$ entering $\sigma^{\rm{i} \rightarrow \rm{d}}_{\rm{neec}}$ in the Lorentz profile.
The factor $P_q$ occurring in Eq.~(\ref{eq:neec-plasma.total}) denotes the probability that an ion of charge state $q$ is present in the plasma. The electron flux $\phi_{\rm{e}}$ in the plasma can be written as the product of the density of states $g(E)$, the Fermi-Dirac distribution $f_{\rm{FD}}(E,T_{\rm{e}})$ for a certain electron temperature $T_e$ and the velocity $v(E)$,
\begin{equation}
\phi_{\rm{e}}(E) = g(E) f_{\rm{FD}}(E,T_{\rm{e}}) v(E)\ .
\label{eq:electron.flux.definition}
\end{equation}
The temperature dependence of $\phi_{\rm{e}}$ is only included in the Fermi-Dirac statistics $f_{\rm{FD}}$. The density of states as well as the velocity are determined by taking the relativistic dispersion relation for the free electrons. 
The chemical potential $\mu_{\rm{e}}$ of the electrons occurring in $f_{\rm{FD}}$ is fixed by adopting the following normalization
\begin{equation}
\int dE \, g(E) f_{\rm{FD}}(E,T_{\rm{e}}) = n_{\rm{e}}\ .
\label{eq:electron.flux.norm}
\end{equation}
Thereby, $n_{\rm{e}}$ represents the number density of free electrons.
By using the definition of the NEEC cross section and assuming that the free electron momentum and the NEEC interaction matrix elements are constant over the width of the Lorentz profile $L_{\rm{d}}(E-E_{\rm{d}})$, Eq.~(\ref{eq:neec-plasma.partial}) can be simplified to
\begin{equation}
\lambda^{q,\alpha_{\rm{d}}}_{\rm{neec}} = \frac{2 \pi^2}{p^2} Y_{\rm{neec}}^{\rm{i} \rightarrow \rm{d}} \Phi_{\rm{e}}^{\rm{res}}(E_{\rm{d}})\ ,
\label{eq:neec-plasma.partial2}
\end{equation}
where the resonant electron flux is defined by
\begin{equation}
\Phi_{\rm{e}}^{\rm{res}}(E_{\rm{d}}) = \int dE \, L_{\rm{d}}(E-E_{\rm{d}}) \phi_{\rm{e}}(E)\ .
\label{eq:res.electron.flux}
\end{equation}

The net NEEC rate $\lambda_{\rm{neec}}$ provided by Eq.~\eqref{eq:neec-plasma.total} is strongly dependent on the available charge states and free electron energies which are both dictated by the plasma conditions. In turn, the latter will not be constant over time as the plasma undergoes expansion. We proceed to formulate a model that takes the spatial expansion of the plasma into account and provides the temporal dynamics of the plasma parameters required to calculate the net NEEC rates in the plasma environment.

\subsection{Plasma expansion \label{hydro}}

In the scenario under investigation, the plasma-formation occurs on the time scale of the XFEL pulse duration ($\sim$ 100~fs) while the plasma expansion 
time is in the range of picoseconds. Accordingly, we neglect the plasma expansion during its formation in the laser-target
interaction. In order to ascertain the effect of the plasma expansion on atomic processes after the interaction of the laser pulse with the target, we model the expansion of the target plasma by a quasi-neutral expansion of spherical  clusters as studied in the context of the intense optical laser pulses interaction with spherical clusters \cite{KrainovS2002,DitmireDRFP1996,Ditmire1997N,Ditmire1997PRL,Ditmire1998PRA,
LeziusDNS1998,MilchbergMP2001,KimAPM2003,PeanoFS2005,Skopalova2010PRL,Hickstein2014PRL}. We follow the analysis of Ref. \cite{KrainovS2002}  to describe the plasma expansion. 
The target plasma is assumed to maintain a uniform (but
decreasing) density throughout the plasma sphere during the
expansion while the electron
temperature decreases with the adiabatic expansion of the plasma,
\begin{equation}
  \frac{3}{2} n_{\rm{e}} V~dT_{\rm{e}} = -P_{\rm{e}}~dV\ ,
  \label{eq:Teexpan}
\end{equation}
where $n_{\rm{e}}$ is the number density of free electrons, and
$V$ is the volume of the plasma with the radius $R$, i.e., $V = 4\pi
R^3 /3$. Here and in the following, we use the Lorentz-Heaviside natural units $\hbar=c=k_{\rm{B}}=\epsilon_0=\mu_0=1$ 
for the plasma modeling part.
The pressure of free electrons $P_{\rm{e}}$ is given by the
ideal gas law, $P_{\rm{e}} = n_{\rm{e}} T_{\rm{e}}$. Therefore,
the time-dependent electron temperature $T_{\rm{e}}(t)$ and the plasma radius
$R(t)$ satisfy the relation
\begin{equation}
  T_{\rm{e}} = T_{\rm{e,0}} \left(\frac{R_0}{R}\right)^2\ ,
  \label{eq:Tetime}
\end{equation}
where $T_{\rm{e,0}}$ is the initial electron temperature and $R_0$
the initial plasma radius. During the plasma expansion, the electrons lose their thermal energy to the ions  resulting into the electron and ion kinetic energies 
\begin{subequations}
\begin{equation}\label{eq:TiTe}
  n_{\rm{i}} \frac{d}{dt} \left( \frac{3}{2} T_{\rm{i}} \right) =
  - n_{\rm{e}} \frac{d}{dt} \left( \frac{3}{2} T_{\rm{e}} \right)\ ,
\end{equation}
\begin{equation} \label{eq:eksurion}
  \frac{1}{2} m_{\rm{i}} \left(\frac{dR}{dt}\right)^2 = \frac{3}{2}
  T_{\rm{i}}\ ,
 \end{equation}
\end{subequations}
with $m_{\rm{i}}$ being the ion mass. The equation of plasma expansion then reads \cite{KrainovS2002}
\begin{equation}
  m_{\rm{i}} \frac{d^2 R}{dt^2} = 3\frac{n_{\rm{e}}}{n_{\rm{i}}}
  \frac{T_{\rm{e,0}}~R_0^2}{R^3} = 3 Z_{\rm{i}} \frac{T_{\rm{e,0}}~R_0^2}{R^3}\ ,
  \label{eq:expanri}
\end{equation}
where $Z_{\rm{i}}\equiv n_{\rm{e}}/n_{\rm{i}}$ is the ratio of
the electron density to the ion density. In the
quasi-neutral limit, $Z_{\rm{i}}$ is therefore the average charge state of the
ions. Integrating once for a fixed $Z_i=Z_0$ yields
\begin{equation}
  \left(\frac{dR}{dt}\right)^2 =
  v_{\rm{s}}^2 \left( 1- \frac{R_{0}^2}{R^2}
  \right),
  \label{eq:vspeed}
\end{equation}
where $v_{\rm{s}} = (3 T_{\rm{e,0}} Z_0 / m_{\rm{i}})^{1/2}$ is the ion sound velocity which, in the limit $R\rightarrow \infty$, is the characteristic speed for the
plasma expansion \cite{KrainovS2002,DitmireDRFP1996}. It may be noted that Peano \emph{et al.}
\cite{PeanoPMCS2006,PeanoCPMS2007} analyzed the expansion of
spherical nanoplasmas with the Vlasov-Poisson equations, the
particle-in-cell (PIC) method, and the ergodic model
\cite{PeanoPMCS2006}. Comparisons between their results
\cite{PeanoPMCS2006,PeanoCPMS2007} and those via the hydrodynamic
expansion [Eq.\eqref{eq:expanri}] (atomic processes  not included) for
 plasmas with the dimensionless parameter  $\tilde{T}_0 = 3 \lambda_{\rm{D}}^2/R_0^2=7.2 \times 10^{-3}$ or $\tilde{T}_0 =
7.2 \times 10^{-2}$, where $\lambda_{\rm{D}}$ denotes the Debye length for the electrons, show that the hydrodynamic expansion model can adequately describe the expansion of spherical
cluster targets heated by an intense optical laser pulse. In addition, 1D and 2D PIC simulations
using the EPOCH code \cite{EPOCH} for plasmas (atomic processes  not included)
with the parameters under consideration in this
article have also been performed. The expansion time obtained by the
hydrodynamic expansion model is in good agreement with the PIC
simulation results (not shown here).

The quasi-neutral hydrodynamic plasma expansion described above requires the plasma density and the charge state as  input parameters. One can exploit the different time scales involved in  the plasma formation and subsequent plasma expansion, and as a first step calculate the average charge state by using the electron temperature from the plasma expansion model. One can then use this charge state to estimate the plasma expansion and consequently the electron temperature at a later time, thus establishing a feedback loop between the effective atomic processes (represented by the average charge state) and the plasma expansion. This is discussed in Sec. \ref{atomic}.

\section{Results and discussion \label{results}}

We investigate and quantify the direct and secondary nuclear excitation in a normal incidence setup for two species of nuclei, (i)  $^{93}$Mo isomer triggering where the isomeric state $IS$ is depopulated via a 4.85~keV transition to an above lying triggering level $T$, as depicted in Fig.~\ref{fig:neec}; and (ii) the resonant driving of the 14.4~keV transition in ${}^{57}$Fe from the ground state $G$ to the first excited level $E$. For both nuclear targets, we assume a thickness of 1 $\mu$m, which for the case of ${}^{57}$Fe corresponds to a resonant thickness of $\xi=3.9$. The laser focal spot is assumed to be 10 $\mu$m$^2$, i.e., focal radius of  approx.~1.8 $\mu$m. 

In the case of ${}^{57}$Fe the initial state is the stable ground state of the isotope, such that bulk iron samples enriched with the ${}^{57}$Fe
isotope (natural abundance 2.2\%) can be fabricated. The $^{93}$Mo solid-state sample should on the other hand contain nuclei in the isomeric state at 2.5 MeV excitation energy. The isomers can be produced in 
$^{93}_{41}$Nb(p,n)$^{93\mathrm{m}}_{\phantom{m} 42}$Mo reactions \cite{exfor},  directly embedded into 1~$\mu$m thick solid-state niobium foils \cite{Jonas2013.PRL}. Based on the production reaction cross section values \cite{exfor} ($\sim30$~mb), we estimate that a ${}^{93\rm m}$Mo isomer density of 10$^{16}$~cm$^{-3}$  can be achieved in the solid-state 1 $\mu$m-thick Nb foils \cite{Jonas2013.PRL-supplement} using standard proton 
beams like the LINAC at GSI \cite{GSI-web,LINAC-web}. 
(Considering possible transport difficulties between isomer production and final experimental sites, it would be of great advantage to have an in situ production of the isomers directly at XFEL facilities, in conjunction with optical-laser-driven proton acceleration.)
The majority of atoms in the isomeric sample belong to the  niobium species ($Z=41$) and for the plasma parameter estimates, their interaction with the XFEL photons will play the dominant role. However, due to the very close atomic number values of molybdenum and niobium, the photoelectric response of the two atomic species is very similar. 

In the following we address in more detail the XFEL parameter choice and present our photoexcitation and NEEC numerical results for the two considered nuclear transitions.

\subsection{Light sources}
The advent and commissioning of the  XFEL  promise significant progress in the field of nuclear quantum optics. It offers in particular the possibility to investigate the direct laser-nucleus interaction with coherent, highly brilliant x-ray pulses. Characteristic for this new kind of radiation sources are the small wavelengths in the x-ray regime, high power, high brilliance and coherent light pulses. The pulse duration lies usually in the range of hundreds of femtoseconds. 

Currently, there are two operating XFEL facilities worldwide, the LCLS \cite{LCLS-web} at SLAC in Stanford and the SACLA \cite{SACLA-web} in Japan. The LCLS provides photons with an energy of approximatively 10~keV with an average spectral brightness of up to $2.7\times10^{22}$ photons/(s mrad$^2$ mm$^2$ 0.1\%bandwidth). The SACLA facility in Japan delivered the so-far highest  photon energy of 19.5~keV. In addition, several XFEL machines are in construction like the European XFEL at DESY in Hamburg, the SwissFEL at the Paul Scherrer Institute in Switzerland, and MaRIE at the Los Alamos National Laboratory in the United States. The European XFEL, for instance, is expected to achieve photon energies up to 24.8~keV. The XFEL generated higher harmonics may even provide photon pulses with energies above 25~keV.

In addition to the brightness of the XFEL radiation, the coherence properties are in particular prominent in comparison to broad-band synchrotron radiation. However, while the XFEL pulses are fully spatially coherent, their temporal coherence is poor because of random fluctuations in the electron charge density at the start-up of the SASE process. Until now, there are two ideas how to tackle the problem of the poor temporal coherence. The first idea is to load the undulator with an already seeded light pulse in order to reduce shot-to-shot fluctuations at the start-up (seeded XFEL) \cite{Feldhaus1997.OC,Saldin2001.NIaMiPRA,DESY-seeding}. The second idea is to construct an x-ray cavity which directs the light pulse several times through the undulator (XFEL oscillator, XFELO) \cite{Kim2008.PRL}.

In this work we consider the parameters of the currently running XFELs, the LCLS and the SACLA, and of the European XFEL still under construction at DESY. Additionally, we will also provide results for the XFELO in order to point out the importance of temporal coherent radiation pulses in the direct laser-nucleus interaction. The corresponding laser parameters are summarized in Table \ref{table:laser}. Here, we assume a moderate laser focusing on a spot of 10~$\mu$m$^2$ (a focal length of 7~nm has for instance already been achieved in Ref. \cite{Mimura2010.NP}).

\renewcommand{\arraystretch}{1.2}
\begin{table}
  \centering
  \begin{tabular}{lcccc}
  \hline\hline
  & & & & \tabularnewline[-0.4cm]
  Parameter & LCLS & SACLA  & Eur.~XFEL  & XFELO  \tabularnewline 
  & & & & \tabularnewline[-0.4cm] \hline
  & & & & \tabularnewline[-0.4cm]
  $E_{\rm{max}}$ (eV) & 10300 & 19600 & 24800 & 25000 \tabularnewline
  $BW$ & 2$\times$10$^{-3}$ & 2.2$\times$10$^{-3}$ & 8$\times$10$^{-4}$ & 1.6$\times$10$^{-7}$ \tabularnewline
  $T_{\rm{pulse}}$ (fs) & 100 & 100 & 100 & 1000 \tabularnewline
  $T_{\rm{coh}}$ (fs) & 2 & --\footnote{In our calculations we assumed 10$\%$ of the pulse duration, i.e., 10~fs.} & 0.2 & 1000 \tabularnewline
  $P_{\rm{peak}}$ (W) & 1.5--4$\times$10$^{10}$ & 10$^{10}$ & 2$\times$10$^{10}$ & 4.1$\times$10$^{9}$ \tabularnewline
  $I_{\rm{peak}}$ ($\frac{\rm{W}}{\rm{cm}^2}$) & 3.9$\times$10$^{17}$ & 9.8$\times$10$^{16}$ & 2.0$\times$10$^{17}$ & 4.0$\times$10$^{16}$ \tabularnewline
  $f_{\rm{rep}}$ (Hz) & 30 & 10 & 2.7$\times$10$^{4}$ & 10$^{6}$ \tabularnewline 
  & & & & \tabularnewline[-0.4cm] \hline\hline
  \end{tabular}
  \caption{The approximate maximal achievable photon energy $E_{\rm{max}}$, bandwidth $BW$, pulse duration $T_{\rm{pulse}}$, coherence time $T_{\rm{coh}}$, peak power $P_{\rm{peak}}$, peak intensity $I_{\rm{peak}}$ and pulse repetition frequency $f_{\rm{rep}}$ for the four considered XFEL facilities: LCLS \cite{LCLS-tdr,Emma2010.NP,Gutt2012.PRL}, SACLA \cite{SACLA-tdr,Ishikawa2012.NP}, Eur.~XFEL \cite{EurXFEL-tdr} and XFELO \cite{Kim2008.PRL}. A focal spot of 10~$\mu$m$^2$ is assumed. }
  \label{table:laser}
\end{table}
\renewcommand{\arraystretch}{1}

A general hindrance in the laser-induced nuclear excitation is that only a small fraction of the laser photons really fulfills the nuclear resonance condition due to the usually small nuclear transition widths. This has two major consequences. First, since only a small number of nuclei is excited per pulse, the repetition rate of the XFEL facility is a key ingredient for an effective driving of nuclear transitions. The repetition rates of the LCLS and the SACLA have with 30~Hz and 10~Hz, respectively, the same order of magnitude. The future European XFEL is expected to provide light pulses with a repetition rate of $2.7\times10^4$~Hz and the XFELO even with a frequency of $10^6$~Hz, which is 3 and respectively 4-5 orders of magnitude higher than the corresponding values of the already operational XFELs. 

The second consequence is that we need to introduce an effective laser intensity which accounts for the mismatch between photon and nuclear transition energy \cite{Palffy2008.PRC}, namely
\begin{equation}
I_{\rm{eff}} = \frac{\Gamma_{\rm{nucl}}}{\Gamma_{\rm{laser}}} I\ .
\label{eq:eff.intensity}
\end{equation}
Thereby, $\Gamma_{\rm{nucl}}$ denotes the nuclear transition width and $\Gamma_{\rm{laser}}$ the bandwidth of the laser pulse. The effective field amplitude $E_{\rm{eff}}$ required in Eq.~(\ref{matrHI}) in order to obtain numerical results is proportional to $\sqrt{I_{\rm{eff}}}$.

\subsection{Direct photoexcitation}

\renewcommand{\arraystretch}{1.2}
\begin{table*}
  \centering
  \begin{tabular}{lcccccccc}
  \hline\hline
  & \hspace*{0.3cm} & & & & \hspace*{0.3cm} & & & \tabularnewline[-0.4cm]
  & & \multicolumn{3}{c}{${}^{93}$Mo} & & \multicolumn{3}{c}{${}^{57}$Fe} \tabularnewline
  & & & & & & & & \tabularnewline[-0.4cm] \cline{3-5}\cline{7-9}
  & & & & & & & & \tabularnewline[-0.4cm]
  & & $I_{\rm{eff}}$ (W/cm$^2$) & $\rho_{\rm{ee}}^{\rm{laser}}$ & $R_{\gamma}^{\rm{laser}}$ (1/s) & & $I_{\rm{eff}}$ (W/cm$^2$) & $\rho_{\rm{ee}}^{\rm{laser}}$ & $R_{\gamma}^{\rm{laser}}$ (1/s) \tabularnewline
  & & & & & & & & \tabularnewline[-0.4cm] \hline
  & & & & & & & & \tabularnewline[-0.4cm]
  LCLS\footnote{\centering In the case of ${}^{57}$Fe, we assumed the x-ray pulse resonant to the 14.4~keV transition, although $E_{\rm{max}}=10.3$~keV for the LCLS.} & & 5.2$\times$10$^{9}$ & 1.9$\times$10$^{-20}$ & 5.6$\times$10$^{-14}$ & & 3.1$\times$10$^{8}$ & 9.1$\times$10$^{-16}$ & 1.9$\times$10$^{-2}$ \tabularnewline
  SACLA & & 1.2$\times$10$^{9}$ & 1.7$\times$10$^{-20}$ & 1.7$\times$10$^{-14}$ & & 7.0$\times$10$^{7}$ & 9.5$\times$10$^{-16}$ & 6.6$\times$10$^{-3}$ \tabularnewline
  Eur.~XFEL & & 6.5$\times$10$^{9}$ & 2.4$\times$10$^{-21}$ & 6.5$\times$10$^{-12}$ & & 3.9$\times$10$^{8}$ & 1.2$\times$10$^{-16}$ & 2.2 \tabularnewline
  XFELO & & 6.7$\times$10$^{12}$ & 4.5$\times$10$^{-14}$ & 4.6$\times$10$^{-3}$ & & 4.0$\times$10$^{11}$ & 2.2$\times$10$^{-9}$ & 1.5$\times$10$^{9}$ \tabularnewline
  & & & & & & & & \tabularnewline[-0.4cm] \hline\hline
  \end{tabular}
  \caption{Excited state occupation number $\rho_{\rm{ee}}^{\rm{laser}}$ and signal photon rate $R_{\gamma}^{\rm{laser}}$ for ${}^{93}$Mo and ${}^{57}$Fe. Both $\rho_{\rm{ee}}^{\rm{laser}}$ and $R_{\gamma}^{\rm{laser}}$ have been calculated by considering only the direct photoexcitation. The laser parameters shown in Table \ref{table:laser} are employed here.}
  \label{table:photoexcitation}
\end{table*}
\renewcommand{\arraystretch}{1}

The photoexcitation rates are calculated following the formalism in Ref.~\cite{Palffy2008.PRC} using reduced transition probability values from Ref.~\cite{ENSDF}. The limited laser coherence time is accounted for by introducing a corresponding decoherence rate in the Lindblad operator. 
The numerical solution of the master equation for the density matrix is then carried out with the procedures implemented in \textit{Mathematica} \cite{mathematica}. We considered realistic parameters for the coherent high-frequency light sources shown in Table \ref{table:laser}. 
Results for the total population of the excited level $\rho_{\rm{ee}}^{\rm{laser}}$  after a single laser pulse are presented in Table \ref{table:photoexcitation}. These populations are obtained by summing the elements $\rho_{\rm{ee}}(M_{\rm{e}})=\langle I_{\rm{e}} M_{\rm{e}}|\rho| I_{\rm{e}} M_{\rm{e}}\rangle$ evaluated at $t=T_{\rm{pulse}}$ over all possible projection quantum numbers $M_{\rm{e}}$. For both  ${}^{93m}$Mo and ${}^{57}$Fe cases, the XFELO delivers the highest excitation rates per pulse among the considered laser facilities.

Comparing the effective laser intensities displayed in Table \ref{table:photoexcitation}, it can be seen that the XFELO provides with 6.7$\times$10$^{12}$~W/cm$^2$ for ${}^{93m}$Mo and 4.0$\times$10$^{11}$~W/cm$^2$ for ${}^{57}$Fe the highest value of $I_{\rm{eff}}$. The effective intensity of the LCLS, for instance, is 3 orders of magnitude smaller. Thus, due to its narrow bandwidth, the XFELO is able to provide in average 1000 times more resonant photons per pulse than the LCLS, SACLA or the Eur.~XFEL. 

Another, possibly less obvious reason for the outstanding excitation properties of the XFELO are its coherence properties. As already mentioned in the previous section, all currently operating XFELs lack a good temporal coherence (indicated by $T_{\rm{coh}}$ in Table \ref{table:photoexcitation}) because of the random fluctuations in the initial electron charge density. The poor temporal coherence is one of the main limiting factors for the nuclear excitation. Considering for instance totally coherent x-ray pulses ($T_{\rm{coh}}=\infty$) increases the nuclear excitation between 4 and 6 orders of magnitude.

Experimentally, the nuclear excitation is best accessible by measuring the number of photons of a specific transition emitted in the decay of the excited nuclear state $E$. In the case of isomer triggering, for instance, the triggering level $T$ first decays to an intermediate state $F$ which then subsequently falls down via a decay cascade to the nuclear ground state $G$. In this cascade a characteristic photon of 1~MeV is emitted which can be used as signature for the isomer triggering. The rate of $\gamma$-ray signal photons at the detector is directly proportional to $\rho_{\rm{ee}}$,
\begin{equation}
R_{\gamma}^{\rm{laser}} = N_0 f_{\rm{rep}} B^{T \to F} \frac{1}{1+\alpha_{\rm{ic}}} \rho_{\rm{ee}}\ ,
\label{eq:signal.rate}
\end{equation}
where  $N_0=n_0 A_{\rm{foc}} L$ represents the number of nuclei present in the focal spot $A_{\rm{foc}}$ and $\alpha_{\rm{ic}}$ stands for the IC coefficient of the transition producing the 1~MeV signal photon. Furthermore, $B^{T \to F}$ denotes the 
branching ratio, i.e,  the probability of a nucleus in $T$ to not fall back to the state $IS$.  For ${}^{93}$Mo, this probability that the triggering level decays to the ground state via the emission of the 1~MeV signal photon can be approximated by one. In the case of ${}^{57}$Fe, the resonantly scattered photons via the transition from the excited state $E$ to the ground state $G$ can be used as detection signal. In this case, a similar expression as Eq.~(\ref{eq:signal.rate}) can be used, with a branching ratio value representing the probability that the excited nuclear state $E$ decays radiatively. Taking the collective channel into account, this probability is approx.~82\% for a sample thickness of 1~$\mu$m ($\xi=3.9$).

Results for $R_{\gamma}^{\rm{laser}}$ are also shown in Table \ref{table:photoexcitation}. Since the nuclear excitation per pulse is typically very small, the pulse repetition frequency $f_{\rm{rep}}$ of the laser plays a crucial role in order to have detectable signal rates. The Eur.~XFEL, for instance, is expected to provide two orders of magnitude larger rates than the LCLS although the excited state population per pulse is smaller, simply due to the high repetition frequency of 27000 pulses per second. The XFELO may  produce $4.6\times10^{-3}$ and $1.5\times10^{9}$ signal photons per second for ${}^{93m}$Mo and ${}^{57}$Fe, respectively. The large difference between ${}^{93m}$Mo and ${}^{57}$Fe comes mainly from the magnitude of the interaction matrix elements and the number of nuclei present in the sample, which are orders of magnitude apart. This difference also compensates that the effective intensity for the resonant photoexcitation of ${}^{57}$Fe is about one order of magnitude smaller than for ${}^{93m}$Mo.

\subsection{Secondary NEEC in a stationary plasma}

The net NEEC rate $\lambda_{\rm{neec}}$ provided by Eq.~\eqref{eq:neec-plasma.total} is strongly dependent on the available charge states and free electron energies which are in turn both dictated by the plasma conditions. The charge state distribution (CSD) can be calculated by applying the collisional-radiative model implemented in FLYCHK \cite{FLYCHK}. In that model the CSD is  completely determined by fixing the electron temperature $T_{\rm{e}}$ and the ion density $n_{\rm{i}}$ present in the plasma.  
In the following we investigate the role of the plasma conditions, considered to be stationary here, on $\lambda_{\rm{neec}}$ for the numerical case of ${}^{93m}$Mo isomer triggering.

The microscopic transition rates $Y_{\rm{neec}}$ given in Eq.~(\ref{eq:neec.rate}) are computed numerically  following the formalism developed in Ref.~\cite{Palffy2007.PRL}. The electronic wave functions and binding energies are calculated by the relativistic multi-configurational Dirac Fock (MCDF) method implemented in the computer code GRASP92 \cite{GRASP}. The probabilities $P_q$ are determined by the FLYCHK-CSD results  and the electron flux $\phi_{\rm{e}}$ is calculated following the expressions in Sec.~\ref{neecinplasma}. Finally, the convolution integral over the Lorentz profile and the electron flux appearing in Eq.~(\ref{eq:neec-plasma.partial}) is solved by approximating the resonance profiles by Dirac delta distributions centered at $E_{\rm{d}}$.

\renewcommand{\arraystretch}{1.2}
\begin{table}
  \centering
  \begin{tabular}{lcc}
  \hline\hline
  & & \tabularnewline[-0.4cm]
  capture & $\Phi^{\rm{res}}_{\rm{e}}$ & $S^{IS \rightarrow F}_{\rm{neec}}$ \tabularnewline
  orbital & [1/m$^2$/s/eV] & [b eV] \tabularnewline 
  & & \tabularnewline[-0.4cm] \hline
  & & \tabularnewline[-0.4cm]
  3$d_{3/2}$ & $1.10\times 10^{31}$ & $3.04\times 10^{-8}$ \tabularnewline
  3$d_{5/2}$ & $1.09\times 10^{31}$ & $4.28\times 10^{-8}$ \tabularnewline
  4$d_{3/2}$ & $2.77\times 10^{30}$ & $1.05\times 10^{-8}$ \tabularnewline
  4$d_{5/2}$ & $2.76\times 10^{30}$ & $1.50\times 10^{-8}$ \tabularnewline
  5$d_{3/2}$ & $1.59\times 10^{30}$ & $5.10\times 10^{-9}$ \tabularnewline
  5$d_{5/2}$ & $1.59\times 10^{30}$ & $7.28\times 10^{-9}$ \tabularnewline
  & & \tabularnewline[-0.4cm] \hline\hline
  \end{tabular}
  \caption{NEEC  case study for a 350~eV plasma with ions initially in the charge state $q=24$. The resonant electron flux ($T_{\rm{e}}=350$~eV) and the NEEC resonance strength values are presented for captures into $d_{3/2}$ and $d_{5/2}$ orbitals.}
  \label{table:cut-off}
\end{table}
\renewcommand{\arraystretch}{1}

\begin{figure}
\centering
\includegraphics[width=1.0\linewidth]{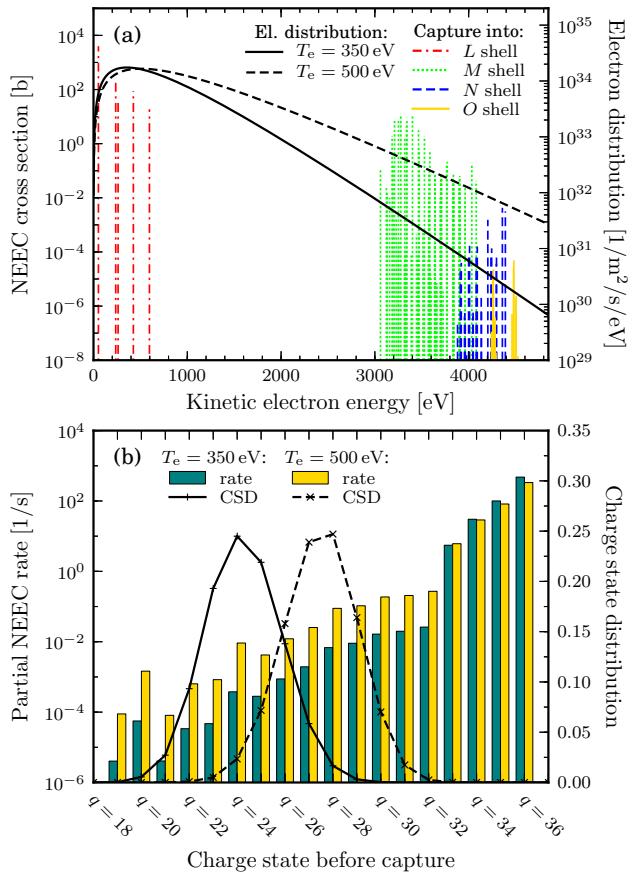}%
\caption{(color online) (a) NEEC resonance cross sections $\sigma_{\rm neec}$ for captures into the $L$, $M$, $N$ and $O$ shell (left axis) together with the electronic energy distribution (right axis). (b) Partial NEEC rate $\lambda^{q}_{\rm neec}$ (left axis) together with the corresponding probabilities of the available CSD (right axis). Results are presented for two plasma temperatures, 350 eV and 500 eV \cite{Jonas2013.PRL}. Reproduced with permission from Phys. Rev. Lett. 112, 082501 (2014). Copyright 2014 American Physical Society.
\label{fig:neec-results}}
\end{figure}

In Fig.~\ref{fig:neec-results}(a) NEEC cross sections are displayed for several capture levels together with the electron flux $\phi_{\rm{e}}$. For the latter we have exemplarily taken the temperatures $T_{\rm{e}}=350$~eV and $T_{\rm{e}}=500$~eV. The figure shows that the NEEC cross section decreases by going to more loosely bound capture levels. Due to the reduction of the binding energy, the kinetic energy of the initially free electron increases in order to still fulfill the NEEC resonance condition. Since $\sigma_{\rm{neec}}^{\rm{i} \rightarrow \rm{d}}$ is inversely proportional to the squared electron momentum, this leads to a decrease of the NEEC cross section. 

Moreover, by comparing the cross section with the electron flux it can be seen that for $T_{\rm{e}}=350$~eV and $T_{\rm{e}}=500$~eV the resonance energies for $M$, $N$ and $O$ shell recombinations are located at the tail of the electron distribution which shrinks exponentially. This behavior is once more explicitly shown in Table \ref{table:cut-off} where the resonant electron flux $\Phi_{\rm{e}}^{\rm{res}}$ for $T_{\rm{e}}=350$~eV and the integrated cross section---the so-called resonance strength $S_{\rm{neec}}$---are presented for increasingly less bound capture shells. We have considered the example of Mo ions with initially fully occupied $K$, $L$, $M_1$, $M_2$ and $M_3$ shells, corresponding to charge state $q=24$. Moreover, NEEC into $d_{3/2}$ and $d_{5/2}$ orbitals has been considered. The results clearly show that both the NEEC resonance strength and the number of resonant electrons decrease by going to higher principal quantum numbers. This behavior can be recovered for all capture channels and initial charge states, as long as the NEEC resonance energy is much larger 
than the plasma temperature. Hence, a ``cut off'' level can be found for each charge state $q$ from which on the NEEC excitation can be neglected. Numerical results for $\lambda_{\rm{neec}}$  shown in the following always employ such cut offs.

For the final NEEC reaction rate calculation, the three main ingredients are
the electron energy distribution in the plasma, the availability of capture charge states and the magnitude of the NEEC cross sections $\sigma_{\rm neec}$. 
In order to maximize the NEEC excitation rate one would prefer to have as many resonant free electrons as possible. As shown in Fig.~\ref{fig:neec-results}(a) this is the case for the resonance energies of NEEC into the $L$ shell which lie in the region around the maximum of $\phi_{\rm{e}}$ in the case of $T_{\rm{e}}=350$~eV and $T_{\rm{e}}=500$~eV. However, as can be seen from the charge state distributions shown in Fig.~\ref{fig:neec-results}(b), the number of available charge states in the plasma will be limited for a given electron temperature $T_{\rm{e}}$. For instance, in the case of $T_{\rm{e}}=350$~eV and $T_{\rm{e}}=500$~eV only Mo charge states up to $q=29$ and $q=32$ are present in the plasma, respectively. Opening the NEEC resonance channels for the $L$ shell requires the Mo ions to be at least in the charge state $q=33$. Hence, we conclude that $L$-shell capture is not possible for the considered temperatures in Fig.~\ref{fig:neec-results}.

In addition to the molybdenum CSD, Fig.~\ref{fig:neec-results}(b) shows also the nuclear excitation probability in dependence of the charge state $q$  obtained from Eq.~\eqref{eq:neec-plasma.partial} via a summation over all contributing $\alpha_{\rm{d}}$. Again the temperatures $T_{\rm{e}}=350$~eV and $T_{\rm{e}}=500$~eV are considered. Our results show that the nuclear excitation probability grows with $q$ because higher charge states open NEEC capture channels more deeply bound to the nucleus. 

Finally, looking at the dependence of $\lambda_{\rm{neec}}$ on the electron temperature $T_{\rm{e}}$ we can conclude that increasing the temperature leads to a higher NEEC excitation rate for mainly two reasons: (i) higher temperatures involve higher available charge states which renders the capture into deeply bound electron shells possible; (ii) higher temperatures lead to an increase of the total number of available free electrons enhancing the resonant electron flux.

\subsection{Initial plasma conditions \label{plasmainit}}
In contrast to optical or infrared laser light, x-rays are able to produce directly inner shell holes by photoionization and penetrate much further into the material leading to a very uniform irradiation of the solid. Moreover, the solid-state target is heated very rapidly by the laser pulse with  duration of about 100~fs. On this time scale the ionic motion is negligible resulting in a plasma near to solid-state density. This rapid, isochoric heating of the plasma (the volume is nearly unaffected during the formation process) enables us to consider uniform temperature and density distributions at short times after the laser pulse.

The laser-induced creation of the plasma is mainly dominated by two processes: (i) photoionization, and, (ii) Auger decay. In the latter process an electron on a higher atomic shell decays to an inner shell hole with the simultaneous emission of another  electron into the continuum.
The process of photoionization prefers the interaction with inner shell electrons provided that the photon energy exceeds the ionization potential of the electrons. In the case of niobium, for instance, the ionization edges of the $K$ and $L_{1}$ shells lie at 19~keV and 2.7~keV \cite{Henke1993.ADaNDT}, respectively. Hence, considering the resonant 4.85~keV x-rays, at most $L$ shell holes can be produced. The laser-produced inner shell holes are then subsequently refilled by either radiative or Auger decay. 

In order to get a rough  estimate of the initial plasma conditions directly after its creation, we follow the procedure used in Ref.~\cite{Tesla-TDR-2001}. We neglect the hydrodynamic expansion and the radiative losses for  the first 100~fs when the laser pulse is present. (We note here that an extension of FLYCHK, the SCFLY code \cite{SCFLY}, was particularly designed to model the plasma generation phase. Unfortunately, SCFLY is not freely available.)
Assuming an instantaneous equilibration time, equations of state (EOS) can be used to estimate the initial plasma conditions. Thereby, the deposited internal energy per pulse can be approximated by $e(J/g) = I_{\rm{peak}}(W/cm^2) \kappa(cm^2/g) T_{\rm{pulse}}(s)$ where $\kappa$ represents the photoabsorption coefficient. Using the EOS tables \cite{MPQeos} $e(T_{\rm{e}},\rho_0)$, the initial plasma temperature immediately after the creation process can be determined. Note that the ionic density is thereby assumed to remain initially at its solid-state density $\rho_0$ and at room temperature.

By using the LCLS laser parameters  presented in Table \ref{table:laser}, we obtain an initial temperature of $T_{\rm{e,0}} \approx 350$~eV for the Nb target. For the calculation we used a photoabsorption coefficient of 551.6 cm$^2$/g. Analogously, the initial stage of the Fe plasma can be estimated. Since due to the higher photon energy, the photoabsorption coefficient (63.0 cm$^2$/g) is much smaller in this case, we obtain a colder electron temperature from start on, namely $T_{\rm{e,0}} \approx 75$~eV.

\subsection{Atomic effects in plasma expansion \label{atomic}}

In dense plasmas, atomic processes such as recombination and ionization
are expected to play an important role on the plasma dynamics. In the following we sketch
an approximate way to include the effects of
the atomic processes on the hydrodynamic plasma expansion model described in Sec.~\ref{hydro} by
using numerical results from the FLYCHK code \cite{FLYCHK}. 

FLYCHK was developed for describing plasma states  like warm dense matter, highly transient states of matter or extremely hot and dense matter. It employs a schematic atomic structure in order to provide a fast and widely applicable plasma diagnosis tool. Each atomic level is represented only by its principal quantum number $n$. The atomic energy levels are computed from ionization potentials where the effect of the electronic continuum depression occurring in plasmas is taken into account by employing the model of Stewart and Pyatt \cite{SPJ66}. The population kinetics model implemented in FLYCHK is based on rate equations including radiative and collisional transitions between bound states, photoionization, collisional ionization processes, Auger decay, electron capture, radiative recombination and three-body recombination. These rate equations are solved for a finite set of atomic levels which consists of ground states, single excited states ($n\le10$), autoionizing doubly excited states and 
inner shell excited states for all possible ionic stages.

With the model briefly described above, FLYCHK is able to determine
ionization and level population distributions of a plasma (for some
given conditions such as a given electron temperature and density).
It can be applied for low-to-high $Z$ ions under most conditions of
laboratory plasmas, in either steady-state or time-dependent
situations \cite{FLYCHK}.

The plasma under consideration here is a cold dense plasma. Considering again the numerical
example of isomer triggering of ${}^{93m}$Mo, the initial
density of niobium ions is considered to be solid density and the initial
electron temperature is several hundred eV. The result from FLYCHK
shows that the time required to reach the steady state (with regard to the
atomic processes) of such a plasma is on the order of
several hundred femtoseconds. With the radius of the considered plasma of around
$1.8~\mu$m (according to the initial condition that the laser spot radius is
around $1.8~\mu$m), the characteristic time scale for
the plasma expansion  is on the order of $10$~ps. Thus, the
time scale for reaching the steady state (with regard to the atomic
processes) is much smaller than the time scale of expansion.
Therefore, we can assume that the steady state with regard to the atomic
processes establishes at each time instant during the expansion.
As a first approximation, we can include the effects of atomic
processes to the hydrodynamic expansion model by estimating the
charge state of each time instant using FLYCHK.

\begin{figure}
\includegraphics[width=0.75\columnwidth]{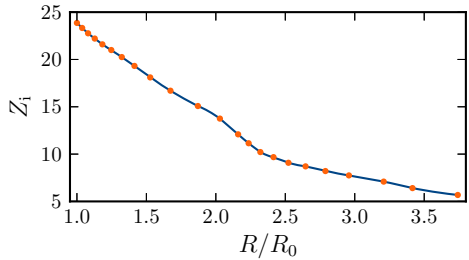}
\caption{(color online). Steady-state average charge of niobium ions obtained
from FLYCHK as a function of $R$ (points) and its interpolation (solid line).
We consider $n_{\rm{i,0}}=5.5\times 10^{22}$~cm$^{-3}$ and
$T_{\rm{e,0}}=350~\rm{eV}$. } \label{fig:zirflychk}
\end{figure}

According to Eq.~(\ref{eq:Tetime}) and to the ion density dynamics given by
\begin{equation}
  n_{\rm{i}} (R) = n_{\rm{i,0}} \left(\frac{R_0}{R}\right)^3\ ,
  \label{eq:nionr}
\end{equation}
we may derive the dependence of the steady-state average charge $Z_{\rm{i}} (R)$ of the plasma on the 
plasma size $R$ using FLYCHK. In the equation above, $n_{\rm{i,0}}$ is the
initial ion number density. Fig.~\ref{fig:zirflychk} shows the
average charge state of the ions for the case of
$n_{\rm{i,0}}=5.5\times 10^{22}$~cm$^{-3}$ and
$T_{\rm{e,0}}=350~\rm{eV}$. The average charge state decreases with reducing
temperature and ion density.

\begin{figure}
\includegraphics[width=0.75\columnwidth]{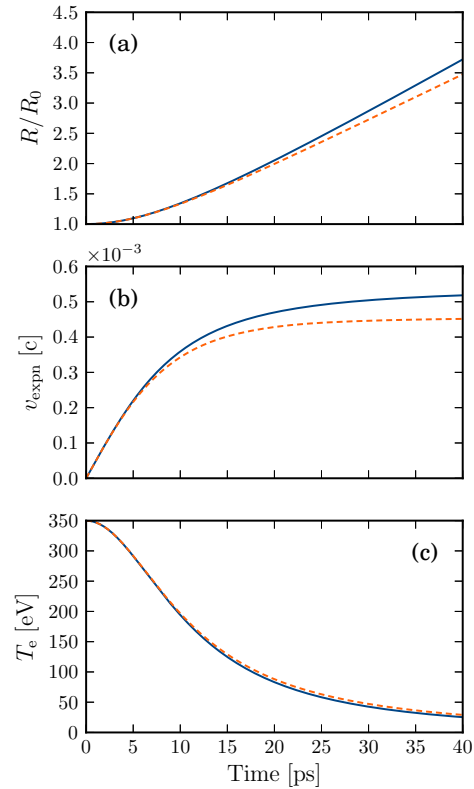}
\caption{(color online). Plasma radius $R$ (a), expansion velocity $v_{\rm{expn}}$ (b), and
electron temperature $T_{\rm{e}}$ (c) as functions of the expansion time. We
use the parameter set 
$n_{\rm{i,0}}=5.5\times 10^{22}$~cm$^{-3}$ (niobium ions) and
$T_{\rm{e,0}}=350~\rm{eV}$. Dashed orange 
curve: the result for the case where the steady-state average charge
$Z_{\rm{i}} (R)$ provided by FLYCHK and shown in Fig.~\ref{fig:zirflychk} is included in the calculations, i.e.,
the atomic processes effects are taken into account. Solid blue
curve: the result for a fixed charge state 
$Z_{\rm{i}} \approx 23.86$.  
}\label{fig:expansion}
\end{figure}

With the average charge state shown in
Fig.~\ref{fig:zirflychk} we can solve Eq.~(\ref{eq:expanri}) in order to
study the plasma expansion. The results are shown in
Figs.~\ref{fig:expansion}. The notation $v_{\rm{expn}} \equiv dR/dt$  has
been introduced. For comparison, the solution of
Eq.~(\ref{eq:expanri}) for the case where the charge state is
fixed to $Z_{\rm{i}} \approx 23.86$ is also presented.  We observe that the
expansion velocity is smaller when the effects of atomic processes in
plasmas are taken into account. This is because the average charge state, originally 
$Z_{\rm{i}} \approx 23.86$, decreases due to atomic processes with the
plasma expansion. Thus, at larger times, the discrepancy between the two
curves in Figs.~\ref{fig:expansion} increases. We note however that the discrepancy
remains small, on the level of 10\%. Due to the plot scale, the discrepancy for $T_{\rm{e}}$ is less visible, although 
it is exactly following the relation (\ref{eq:Tetime}) and it reaches the level of 10\% at later times.

In order to ensure the self-consistency
of our approximation (i.e., that the steady state with regard to the
atomic processes establishes at each time instant during the
expansion), a time-dependent FLYCHK calculation has also been performed, using the time-dependent $T_{\rm{e}}$ shown in
Fig.~\ref{fig:expansion}(c), the time-dependent ion density given by
Eq.~(\ref{eq:nionr}) and $R(t)$ in Fig.~\ref{fig:expansion}(a) as 
input parameters. The result shows that $Z_{\rm{i}}$ from the time-dependent FLYCHK
calculation only slightly deviates ($\sim 5\%$) from the results in
Fig.~\ref{fig:zirflychk} for large $R$, while the results agree well
for small $R$. We conclude that on the degree of accuracy required for our calculations, the approximation used performs sufficiently well.

\begin{figure}
\includegraphics[width=0.75\columnwidth]{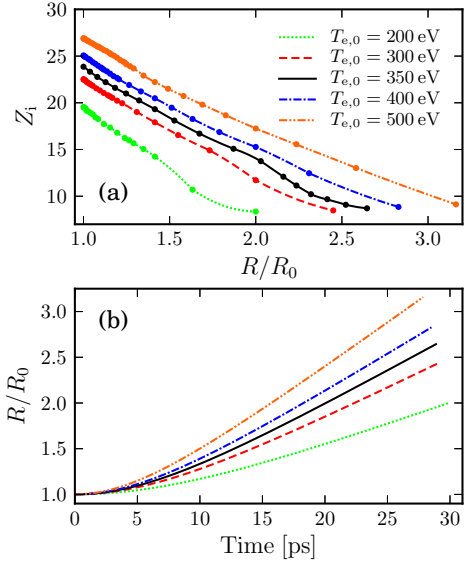}
\caption{(color online). Steady-state average charge of niobium ions (points) obtained
from FLYCHK (a), and plasma radius $R$ as a function of the expansion time (b),
for initial temperatures $T_{\rm{e,0}}=200~\rm{eV}$,
$T_{\rm{e,0}}=300~\rm{eV}$, $T_{\rm{e,0}}=350~\rm{eV}$,
$T_{\rm{e,0}}=400~\rm{eV}$, and $T_{\rm{e,0}}=500~\rm{eV}$. Numerical interpolations are shown by the curves.
We use $n_{\rm{i,0}}=5.5\times 10^{22}$~cm$^{-3}$. The lowest charge state points correspond to cooling down the plasma to $50~\rm{eV}$ electron temperature.} \label{fig:zirte}
\end{figure}

\begin{figure}
\includegraphics[width=0.75\columnwidth]{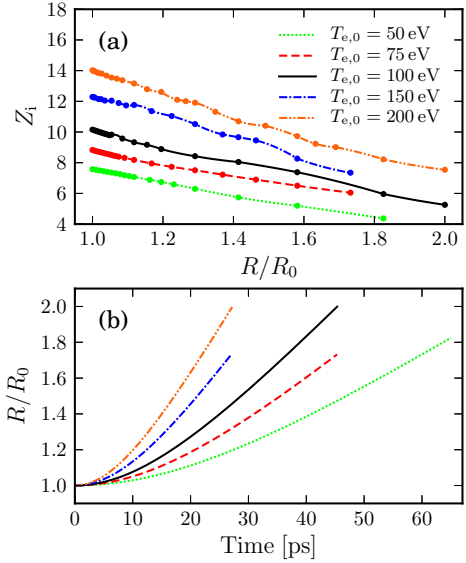}
\caption{(color online). Steady-state average charge of ${}^{57}$Fe ions (points) obtained
from FLYCHK (a), and plasma radius $R$ as a function of the expansion time (b), for initial temperatures $T_{\rm{e,0}}=50~\rm{eV}$,
$T_{\rm{e,0}}=75~\rm{eV}$, $T_{\rm{e,0}}=100~\rm{eV}$,
$T_{\rm{e,0}}=150~\rm{eV}$, and $T_{\rm{e,0}}=200~\rm{eV}$. Numerical interpolations are shown by the curves.
The value
$n_{\rm{i,0}}=8.5\times 10^{22}$ cm$^{-3}$ was used. The end points of the curves 
correspond to the
electron temperature cooled down to $15~\rm{eV}$ for
$T_{\rm{e,0}}=50~\rm{eV}$, $25~\rm{eV}$ for
$T_{\rm{e,0}}=75~\rm{eV}$ and $T_{\rm{e,0}}=100~\rm{eV}$, and
$50~\rm{eV}$ for $T_{\rm{e,0}}=150~\rm{eV}$ and
$T_{\rm{e,0}}=200~\rm{eV}$, respectively.} \label{fig:ziFete}
\end{figure}

We further study the behavior of the average charge state and the plasma radius for several initial temperature values for both niobium (Figs.~\ref{fig:zirte}) and ${}^{57}$Fe (Figs.~\ref{fig:ziFete}) plasmas. Higher initial temperatures lead to higher charge states, with a very similar qualitative behavior at later times in the expansion when the temperature cools down. 


\subsection{Total NEEC excitation}

\begin{figure}
\centering
\includegraphics[width=1.0\linewidth]{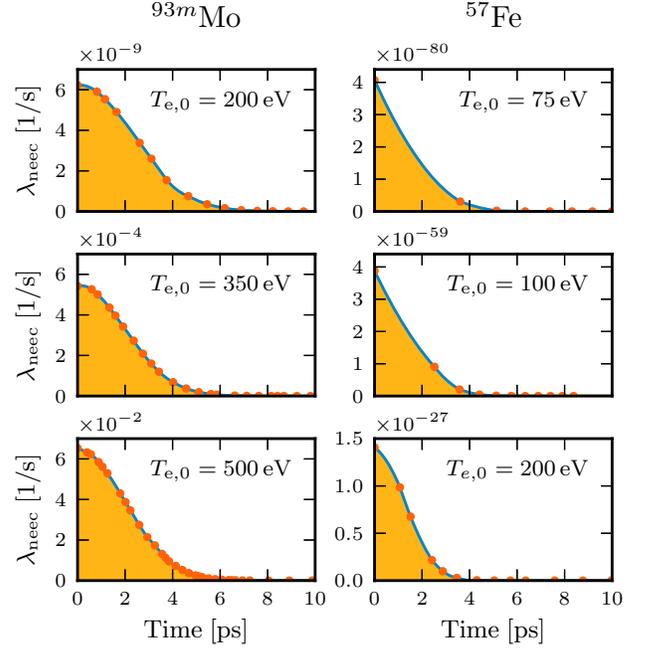}%
\caption{(color online). NEEC reaction rate as a function of time for ${}^{93m}$Mo (left graph column) and ${}^{57}$Fe (right graph column). In the case of ${}^{93m}$Mo  initial electron temperatures of 200~eV, 350~eV and 500~eV have been considered; for ${}^{57}$Fe, the corresponding values are  75~eV, 100~eV and 200~eV, respectively.
\label{fig:neec-hydro}}
\end{figure}

Due to the hydrodynamic expansion of the plasma, the electron temperature and density decrease with time during the plasma expansion leading to a time-dependent net NEEC rate. Fig.~\ref{fig:neec-hydro} presents the time dependence of $\lambda_{\rm{neec}}$ for several initial electron temperatures $T_{\rm{e,0}}$. The ion density is always assumed to be at its solid-state value immediately after the plasma creation. As expected, in all cases $\lambda_{\rm{neec}}$ drops down to zero with time, i.e., with the plasma expansion. With increasing initial temperature, $\lambda_{\rm{neec}}$ takes larger values.

The left column of Fig.~\ref{fig:neec-hydro}  presents  the NEEC reaction rates $\lambda_{\rm{neec}}$ for the case of ${}^{93m}$Mo triggering. Comparing the orders of magnitude of $\lambda_{\rm{neec}}$ it can be seen that the NEEC excitation is strongly dependent on the initial plasma conditions dictated by the laser parameters. Going from $T_{\rm{e,0}} = 200$~eV to $T_{\rm{e,0}} = 500$~eV enhances the NEEC excitation by approximately 7 orders of magnitude. The integration of $\lambda_{\rm{neec}}$ over time provides the excited state occupation number per nucleus (comparable to $\rho_{\rm{ee}}^{\rm{laser}}$),
\begin{equation}
\rho_{\rm{ee}}^{\rm{neec}} = \int_0^{\infty} dt \, \lambda_{\rm{neec}}\ .
\end{equation}
Performing the integration for the case of ${}^{93m}$Mo results in $\rho_{\rm{ee}}^{\rm{neec}} \approx 1.8 \times 10^{-20}$ for $T_{\rm{e,0}} = 200$~eV, $\rho_{\rm{ee}}^{\rm{neec}} \approx 1.4 \times 10^{-15}$ for $T_{\rm{e,0}} = 350$~eV and $\rho_{\rm{ee}}^{\rm{neec}} \approx 1.6 \cdot 10^{-13}$ for $T_{\rm{e,0}} = 500$~eV. For all the three cases, the integration converged after approximately 6~ps. This value gives us the dominant NEEC interaction time during the hydrodynamic plasma expansion. We recall that in Ref.~\cite{Jonas2013.PRL}, the plasma conditions were considered constant for 100 ps. Our present results show that this is not the case and significant changes in the plasma affect the magnitude of the secondary nuclear excitation after already few ps. 

The signal photon rates corresponding to the NEEC-induced triggering of ${}^{93m}$Mo can be obtained by replacing $\rho_{\rm{ee}}$ with $\rho_{\rm{ee}}^{\rm{neec}}$ in Eq.~(\ref{eq:signal.rate}). Assuming LCLS laser parameters, the signal rates 
evaluate to $R_{\gamma}^{\rm{neec}} \approx 5.5 \times 10^{-14}$~s$^{-1}$ for $T_{\rm{e,0}} = 200$~eV, $R_{\gamma}^{\rm{neec}} \approx 4.3 \times 10^{-9}$~s$^{-1}$ for $T_{\rm{e,0}} = 350$~eV and $R_{\gamma}^{\rm{neec}} \approx 4.9 \times 10^{-7}$~s$^{-1}$ for $T_{\rm{e,0}} = 500$~eV. In order to fully exploit the repetition rate of the XFEL, a tape-station system must be used. Moreover, instead of having a single niobium target, a stack of several foils can be employed. This is possible as long as the transmitted laser intensity behind each foil is still strong enough to allow for the plasma formation. In this way an increase in the signal rates of approximatively one order of magnitude may be achievable. 
Nevertheless, the isomer density of $10^{16}$~cm$^{-3}$ remains the main limitation for obtaining higher count rates. 

We turn now to the NEEC reaction rates for  ${}^{57}$Fe presented in the right column of Fig.~\ref{fig:neec-hydro}. While the relevant time scales are similar to the ones of  ${}^{93m}$Mo, on the order of 6 ps, the magnitude of the NEEC rates is in comparison  tens of orders of magnitude below, and for all practical purposes negligible. Let us investigate the reasons for such a dramatic difference. Iron has $Z=26$ and correspondingly smaller binding energies than molybdenum ($Z=42$), while the nuclear excitation energy is larger, 14.4 keV vs. 4.8 keV. The typical capture orbitals for both types of ions lie in the $M$ atomic shell. Correspondingly, the continuum electron energies required for NEEC will be much larger for  ${}^{57}$Fe than for ${}^{93m}$Mo. While the $M$-shell NEEC requires continuum electron energies between 3 and 4 keV, where the electron flux still has large values, in the case of ${}^{57}$Fe the resonant continuum electrons should have more than 13 keV. For this energy value, the plasma at 75 eV temperature provides virtually no electrons at all, leading to the infinitesimal NEEC rate values in Fig.~\ref{fig:neec-hydro}.

\subsection{Comparison between direct and secondary excitation}

In order to obtain comparable results for $\rho_{\rm{ee}}^{\rm{laser}}$ and $\rho_{\rm{ee}}^{\rm{neec}}$ we fix the initial plasma conditions according to the chosen laser parameters. In particular, the laser intensity plays here an important role as outlined in Sec.~\ref{plasmainit}. Restricting ourselves to the LCLS laser parameters, an initial electron temperature of $T_{\rm{e,0}} = 350$~eV is estimated for the niobium target doped with ${}^{93m}$Mo isomers. Comparing the corresponding excited state occupation numbers $\rho_{\rm{ee}}^{\rm{laser}}$ and $\rho_{\rm{ee}}^{\rm{neec}}$, we conclude that the indirect NEEC process is still about five orders of magnitude more efficient than the direct photoexcitation. One important aspect is here that the time of interaction is about 60 times longer for NEEC than for the direct photoexcitation. In contrast to ${}^{93m}$Mo, the indirect NEEC process plays no role for the excitation of ${}^{57}$Fe, since the NEEC reaction rate values are for all practical purposes negligible. 

One important remark here is that the direct photoexcitation can be switched off by tuning the laser off resonance. In contrast, the secondary excitation remains present since the XFEL-produced plasma is not sensitive to small detunings. In the case of ${}^{93m}$Mo for instance, the nuclear transition energy is only known up to an uncertainty of 80~eV. Since the XFEL bandwidths are  several eV, this uncertainty can be a limiting factor for the experimental implementation of the direct photoexcitation channel. The secondary NEEC process occurs instead in a plasma environment with a broad electron distribution where many resonance channels can contribute. Hence, it is much more robust than photoexcitation against such uncertainties in the nuclear parameters. We also note that since the electron fluxes in the plasma are at the studied plasma temperatures several orders of magnitude larger than the corresponding photon fluxes, we neglect in our discussion the possible secondary nuclear photoexcitation process in the plasma. Other studies \cite{Gosselin2004.PRC,Morel2004.PRA,Gosselin2007.PRC,Morel2010.PRC,Stephan2014.PRC} on hot astrophysical plasmas show that this may change at higher, keV plasma temperatures. 
The study of NEEC in plasmas was so far restricted to astrophysical environments \cite{Gosselin2004.PRC,Morel2004.PRA,Gosselin2007.PRC,Morel2010.PRC} or optical-laser-generated plasmas \cite{Harston1999.PRC} where no equivalent of the direct photoexcitation channel under investigation here exists.

The values for $\rho_{\rm{ee}}^{\rm{neec}}$ are strongly dependent on the initial plasma conditions. Theoretically, the initial plasma conditions and in particular the electron temperature should be controllable by changing the laser intensity. Since $\rho_{\rm{ee}}^{\rm{neec}}$ grows exponentially with $T_{\rm{e,0}}$ saturating for large $T_{\rm{e,0}}$ in the considered temperature range and $T_{\rm{e,0}}$ furthermore depends linearly on the laser intensity $I$ between $10^{17}$ and $10^{18}$~W/cm$^2$, we expect for the considered parameter range an exponential dependence of the NEEC-induced excited state population $\rho_{\rm{ee}}^{\rm{neec}}$ on $I$. We note that for this estimate we consider the same initial size for the plasma, i.e., a constant laser focal spot. The direct photoexcitation in contrast follows $\rho_{\rm{ee}}^{\rm{laser}}\propto I$. In addition, the XFEL intensity and consequently the plasma temperature can highly fluctuate from shot to shot. It is therefore reasonable to look at initial temperatures varying around the estimated value as it is realized in Fig.~\ref{fig:neec-hydro}. For ${}^{93m}$Mo, the indirect process is still of the same order of magnitude as the direct one for $T_{\rm{e,0}} = 200$~eV. For an 
initial temperature of 500~eV the indirect NEEC process becomes even competitive with the photo-induced excitation via the coherent XFELO $\rho_{\rm{ee}}^{\rm{laser}} \approx 4.5\times 10^{-14}$ although for the latter a 10 times longer pulse duration than for the LCLS has been considered. In the case of ${}^{57}$Fe, the indirect excitation channel remains negligibly small in comparison to the direct process even by considering an initial electron temperature of 200~eV.

The reason why the nuclear excitation of ${}^{93m}$Mo and ${}^{57}$Fe behave inversely with respect to the direct and indirect excitations is two-fold. First, in the case of ${}^{93m}$Mo triggering the microscopic NEEC cross section is  much larger than the one for resonant photoexcitation. This situation  is reversed in the case of ${}^{57}$Fe. The second and most important reason is related to the plasma conditions. In the case of resonant ${}^{93m}$Mo triggering, the photoabsorption of the 4.85~keV photons by the niobium target ($\kappa=551.6$~cm$^2$/g) is much stronger than the analogon for 14.4~keV photons in iron ($\kappa=63.0$~cm$^2$/g). Naturally, the more photons are absorbed, the more energy is initially deposited in the target leading to a higher initial electron temperature. Then, for NEEC it is important that the available charge states in the plasma and the energy of the free electrons are not that far away from the resonance condition. While this is the case for ${}^{93m}$Mo with a 4.85 keV nuclear transition energy,  it is especially hard to realize in the limit of high nuclear transition energies like for ${}^{57}$Fe.

Let us generalize our results based on the arguments presented above.  A first insight into the relation between the resonant photoexcitation and microscopic NEEC cross sections can be gathered from the corresponding IC coefficients which are defined as the ratio between the IC and radiative decay rates. For high values of $\alpha_{\rm{ic}}$ we expect NEEC to dominate over photoexcitation, which is typically the case for small nuclear transition energies in the keV region. Next, the prevailing initial free electron conditions are strongly dependent on the occurring x-ray atomic photoabsorption. The photoabsorption coefficient grows with increasing $Z$ of the material and decreasing photon energy. Moreover,  shell edges may lead to a stepwise enhancement of $\kappa$. However, a high electron temperature simultaneously leads to higher available charge states which again require smaller kinetic electron energies in order to fulfill the NEEC resonance condition. These opposite trends need to 
be balanced in order to match the available electron energies with the NEEC resonance condition for the open capture channels. Typically, in the temperature range of hundreds of eV the secondary excitation contributes significantly only for transition energies below few  keV.

\renewcommand{\arraystretch}{1.2}
\begin{table*}
  \centering
  \begin{tabular}{lrcrcrrcrrrr}
  \hline\hline
  & & & & & & & & & \tabularnewline[-0.4cm]
  Nuclide & \multicolumn{1}{c}{$E_{\rm{n}}$} & $\mathcal{E}/\mathcal{M}L$ & \multicolumn{1}{c}{$\kappa$} & ion.~shell & \multicolumn{1}{c}{$T_{\rm{e,0}}$} & \multicolumn{1}{c}{$Z_{\rm{i}}$} & capture & \multicolumn{1}{c}{$E_{\rm{ion}}$}  & \multicolumn{3}{c}{$\alpha_{\rm{ic}}$} \tabularnewline
  & \multicolumn{1}{c}{[keV]} & & \multicolumn{1}{c}{[cm$^2$/g]} & & \multicolumn{1}{c}{[eV]} & & shell & \multicolumn{1}{c}{[keV]} & \multicolumn{1}{c}{E1} & \multicolumn{1}{c}{M1} & \multicolumn{1}{c}{E2} \tabularnewline
  & & & & & & & & & \tabularnewline[-0.4cm] \hline
  & & & & & & & & & \tabularnewline[-0.4cm]
  ${}^{201}_{80}$Hg & 1.565 & M1 & 1975 & $N_1$ & 670 & 43 & $N_4$ & 0.378 & & 3.5e+01 & \tabularnewline
  ${}^{193}_{78}$Pt & 1.642 & M1 & 1628 & $N_1$ & 510 & 40 & $N_3$ & 0.519 & & 1.2e+02  \tabularnewline
  ${}^{205}_{82}$Pb & 2.329 & E2 & 911 & $N_1$ & 790 & 45 & $N_3$ & 0.645 & & & 1.8e+07 \tabularnewline
  ${}^{151}_{62}$Sm & 4.821 & M1 & 362 & $M_1$ & 740 & 42 & $M_3$ & 1.420 & & 6.2e+00 &  \tabularnewline
  ${}^{93m}_{42}$Mo & 4.850 & E2 & 552 & $L_1$ & 350 & 24 & $M_2$ & 0.410 & & & 2.5e+04  \tabularnewline
  ${}^{171}_{69}$Tm & 5.036 & M1 & 442 & $M_1$ & 850 & 46 & $M_4$ & 1.515 & & 1.6e+00 &   \tabularnewline
  ${}^{83}_{37}$Rb & 5.236 & M1 & 328 & $L_1$ & 1080 & 32 & $L_1$ & 2.065 & & 7.5e+01 &  \tabularnewline
  ${}^{181}_{73}$Ta & 6.237 & E1 & 300 & $M_1$ & 520 & 36 & $N_2$ & 0.465 & 7.7e--01 & &  \tabularnewline
  ${}^{169}_{69}$Tm & 8.410 & M1 & 118 & $M_1$ & 630 & 43 & $M_5$ & 1.468 & & 1.7e--01 &  \tabularnewline
  ${}^{187}_{76}$Os & 9.756 & M1 & 107 & $M_1$ & 510 & 37 & $N_3$ & 0.468 & & 4.8e--01 &   \tabularnewline
  ${}^{167}_{69}$Tm & 10.400 & M1 & 291 & $L_1$ & 220 & 25 & $N_4$ & 0.180 & & 3.1e--02 &  \tabularnewline
  ${}^{137}_{57}$La & 10.590 & M1 & 166 & $L_1$ & 580 & 37 & $M_3$ & 1.123 & & 3.7e--01 \tabularnewline
  ${}^{134}_{55}$Cs & 11.244 & M1 & 129 & $L_1$ & 1420 & 43 & $L_3$ & 5.012 & & 1.2e+00 &   \tabularnewline
  ${}^{73}_{32}$Ge & 13.285 & E2 & 124 & $K$ & 220 & 18 & $L_3$ & 1.217 & & & 4.1e+02  \tabularnewline
  ${}^{57}_{26}$Fe & 14.413 & M1 & 63 & $K$ & 75 & 12 & $M_1$ & 0.093 & & 1.0e--01 & \tabularnewline
  ${}^{149}_{62}$Sm & 22.507 & M1 & 28 & $L_1$ & 350 & 28 & $M_5$ & 1.080 & & 3.1e--03 &   \tabularnewline
  ${}^{119}_{50}$Sn & 23.871 & M1 & 12 & $L_1$ & 120 & 14 & $N_2$ & 0.089 & & 1.1e--02 &   \tabularnewline
  & & & & & & & & & \tabularnewline[-0.4cm] \hline\hline
  \end{tabular}
  \caption{Low-lying nuclear transitions suitable for resonant photoexcitation via XFELs. Transition energies $E_{\rm{n}}$ and dominating multipolarities $\mathcal{E}/\mathcal{M}L$, photoabsorption coefficients $\kappa$, deepest ionization subshells, estimates of the initial plasma temperature $T_{\rm{e,0}}$, the most probable charge states $Z_{\rm{i}}$ in the plasma, the  most advantageous capture orbitals for NEEC available in the plasma together with the corresponding  ionization energies $E_{\rm{ion}}$  and 
IC coefficients $\alpha_{\rm{ic}}$, respectively, are presented in the columns. The list is sorted by increasing nuclear transition energies  $E_{\rm{n}}$.}
  \label{table:general}
\end{table*}
\renewcommand{\arraystretch}{1}

More concretely, in Table \ref{table:general} we present the list of low-lying nuclear transitions first collected in Ref.~\cite{Junker2012.NJP} with the corresponding transition energy, multipolarity  and IC coefficient $\alpha_{\rm{ic}}$. Also provided are the photoabsorption coefficient $\kappa$, the deepest ionization shell, an approximation of the initial plasma temperature $T_{\rm{e,0}}$, the most probable charge state at this temperature  $Z_{\rm{i}}$, and finally the most NEEC-advantageous available capture shell in the plasma. The latter two quantities have been calculated with FLYCHK. For orientation, also the approximate ionization potential of the most advantageous available capture shell is presented. Since FLYCHK calculations are limited to the range $Z<80$, the entries $Z_{\rm{i}}$ and the capture shell for ${}^{201}_{80}$Hg and ${}^{205}_{82}$Pb have been estimated by taking $Z=79$. The initial temperatures in the table, with the exception of ${}^{93m}$Mo and ${}^{57}$Fe, discussed in detail in Sec.~\ref{plasmainit}, have been obtained  by accounting for the laser energy deposited into the sample with the help of mass photoabsorption coefficients \cite{XCOM}. Due to unavailable EOS tables, the averaged electron temperatures have been then roughly estimated by further considering energy conservation of the inner shell photoionization and the first sequence of Auger decays.

Based on the present analysis, we conclude that isotopes with high nuclear transition energies like ${}^{57}_{26}$Fe, ${}^{149}_{62}$Sm and ${}^{119}_{50}$Sn, which in addition only allow for NEEC into the $M$  or even higher shells, will not present significant secondary nuclear excitation in the XFEL-produced plasma. Nuclei with small transition energies as it is   the case for the first table entries down to
${}^{83}_{37}$Rb are very likely to present significant secondary nuclear excitation in the plasma, due to the abundant electron flux at the required energies and the encouraging IC coefficient. As for the intermediate region with 6 keV$< E_{\rm{n}}<$13 keV, here a more careful analysis is required. The capture into $L$-shell orbitals in the case of ${}^{134}_{55}$Cs and ${}^{73}_{32}$Ge speaks for more available free electrons in the plasma at the resonance energy. However, due to the low $Z$ of germanium, it is likely that only in the case of  ${}^{134}_{55}$Cs secondary NEEC plays any role in the net nuclear excitation. In the cases of ${}^{169}_{69}$Tm, ${}^{187}_{76}$Os, ${}^{167}_{69}$Tm and ${}^{137}_{57}$La, the rather large transition energies, together with $M$- and $N$-shell capture orbitals and the small IC coefficients indicate unlikely strong NEEC influence. Finally, the small IC coefficient in the case of ${}^{181}_{73}$Ta also rather speaks against strong secondary NEEC. However, since the arguments above do not relate to the time scale difference between NEEC and photoexcitation and the unavoidable detuning for the photoexcitation channel, we conclude that dedicated simulations are required to draw solid conclusions for the nuclei in the intermediate energy region.

\section{Summary and Conclusions}

The resonant driving of nuclear transitions with the XFEL is most likely to employ high-density,
solid-state nuclear targets in order to improve the excitation rates. However, in this case
secondary nuclear excitation in the produced cold, dense plasma might occur. We have quantified here
the magnitude of the secondary excitation via NEEC taking into account the plasma dynamics after
the laser pulse. We find that while photoexcitation can only occur during the pulse duration, on the
order of 100 fs, the plasma effects last for several ps until plasma expansion diminishes completely the NEEC rate.
In order to describe the time-dependence of the plasma parameters we employ a hydrodynamical model for the
plasma expansion together with atomic processes input which determines the ion charge distribution at each time instant.
Plasma parameters such as initial temperature, charge state distribution or electron flux play a crucial role for the magnitude of the occurring secondary 
excitation.

Our results show that for small nuclear transition energies and advantageous free electron energy distributions, as it is the case for isomer triggering of  ${}^{93m}$Mo, 
secondary NEEC may exceed by as much as five orders of magnitude the direct photoexcitation by the XFEL. Our more realistic plasma condition analysis thus confirms qualitatively the results
in Ref.~\cite{Jonas2013.PRL}. 
Nevertheless, the signal photon rates for the ${}^{93m}$Mo activation appear to remain below experimental detection, despite several possibilities to increase the triggering efficiency. In addition to enhancing the number of irradiated isomers, one could envisage choosing XFEL parameters such that the conditions for NEEC triggering are optimized.  
The frequency of the XFEL radiation, for instance, has been so far chosen resonant to the nuclear transition under investigation, in order to allow for direct photoexcitation. However, by fully exploiting the theoretical framework derived in this work, 
an optimized parameter set  for more efficient secondary  NEEC-induced triggering of ${}^{93m}$Mo may be deduced.

On the other hand, we also show that for larger nuclear transition energies, in the case of the 14.4 keV M\"ossbauer transition in ${}^{57}$Fe, secondary NEEC can be safely neglected. We may conclude that nuclear quantum optics experiments with bulk isotopically enriched ${}^{57}$Fe samples in normal incidence will not suffer from strong decoherence rates due to plasma-related processes. Caution is however advised for extending these conclusions for experiments with x-ray thin-film planar cavities containing ${}^{57}$Fe iron layers. In these setups which operate in grazing incidence, the working principle of the x-ray cavities is particularly sensitive to radiation damage. Based on our present results, a general set of criteria for identifying the parameter regime for which secondary effects in the plasma are important is presented. We anticipate that our findings will be of relevance for the layout of first nuclear excitation experiments at XFEL facilities in the near future.

\begin{acknowledgments}

The authors are indebted to R. Ramis for providing them with EOS tables for iron, niobium and molybdenum. 

\end{acknowledgments}


\bibliography{mybib-Jonas}

\begin{thebibliography}{97}
\bibitem [{EPO()}]{EPOCH}%
  \BibitemOpen
  \href@noop {} {}\bibinfo {howpublished} {The {EPOCH} code was developed as
  part of the {UK EPSRC} funded projects {EP/G054940/1}}\BibitemShut {NoStop}%
\end{thebibliography}

\end{document}